\documentclass[reprint, amsmath,amssymb, aps, graphicx,superscriptaddress,nofootinbib]{revtex4-2}
\usepackage[utf8]{inputenc}
\usepackage{times}
\usepackage{amsmath}
\usepackage{braket}
\usepackage{hyperref}
\usepackage{rotating}
\usepackage{amsmath}
\usepackage[margin=1.95cm]{geometry}
\usepackage{xfrac}
\usepackage{xcolor}
\usepackage{graphicx}
\usepackage{physics}
\usepackage{lipsum}
\usepackage[title]{appendix}
\usepackage[inline]{enumitem}

\begin{document}
	
\newcommand{\up}{\ket{\uparrow}}
\newcommand{\down}{\ket{\downarrow}}

\newcommand{\saa}{\sqrt{\alpha_A}}
\newcommand{\sab}{\sqrt{\alpha_B}}
\newcommand{\smaa}{\sqrt{1-\alpha_A}}
\newcommand{\smab}{\sqrt{1-\alpha_B}}

\newcommand{\eia}{e^{-i\vartheta_A}}
\newcommand{\eib}{e^{-i\vartheta_B}}

\newcommand{\sea}{\sqrt{\eta_A}}
\newcommand{\seb}{\sqrt{\eta_B}}
\newcommand{\smea}{\sqrt{1-\eta_A}}
\newcommand{\smeb}{\sqrt{1-\eta_B}}

\newcommand{\zea}{\zeta_A}
\newcommand{\zeaa}{\zeta_{AA}}
\newcommand{\zeb}{\zeta_B}
\newcommand{\zebb}{\zeta_{BB}}
\newcommand{\zei}{\zeta_i}
\newcommand{\zeii}{\zeta_{ii}}

\newcommand{\afa}{\hat{a}^{\dagger}_{1,A}}
\newcommand{\afb}{\hat{a}^{\dagger}_{1,B}}
\newcommand{\asa}{\hat{a}^{\dagger}_{2,A}}
\newcommand{\asb}{\hat{a}^{\dagger}_{2,B}}
\newcommand{\afi}{\hat{a}^{\dagger}_{1,i}}
\newcommand{\asi}{\hat{a}^{\dagger}_{2,i}}

\newcommand{\afal}{\hat{a}^{\dagger}_{1,A,r}}
\newcommand{\afbl}{\hat{a}^{\dagger}_{1,B,r}}
\newcommand{\asal}{\hat{a}^{\dagger}_{2,A,r}}
\newcommand{\asbl}{\hat{a}^{\dagger}_{2,B,r}}

\newcommand{\kza}{\ket{0}_A}
\newcommand{\koa}{\ket{1}_A}
\newcommand{\kzb}{\ket{0}_B}
\newcommand{\kob}{\ket{1}_B}

\newcommand{\kzz}{\ket{00}_{AB}}
\newcommand{\kzo}{\ket{01}_{AB}}
\newcommand{\koz}{\ket{10}_{AB}}
\newcommand{\koo}{\ket{11}_{AB}}

\newcommand{\vac}{\ket{0}_\gamma}

\newcommand{\sst}{\sin^2\frac{\theta}{2}}
\newcommand{\cst}{\cos^2\frac{\theta}{2}}

\newcommand{\johs}[1]{\textcolor{blue}{#1}}
	
\newcommand{\figref}[1]{Figure \ref{fig:#1}}
\newcommand{\figlet}[1]{\textbf{(#1)}}
\newcommand{\subfigref}[2]{Figure \ref{#1}#2}

\newcommand{\secref}[1]{Section \ref{#1}}

\newcommand{\eref}[1]{Equation (\ref{eq:#1})}

	\title{
	Entangling remote qubits using the single-photon protocol: an in-depth theoretical and experimental study}
	\author {S. L. N. Hermans}
	\affiliation{QuTech and Kavli Institute of Nanoscience, Delft University of Technology, Delft, The Netherlands}
	\affiliation{Present address: Institute for Quantum Information and Matter, California Institute of Technology, Pasadena, California, USA}
	\author {M. Pompili}
	\affiliation{QuTech and Kavli Institute of Nanoscience, Delft University of Technology, Delft, The Netherlands}
	\affiliation{Present address: Pritzker School of Molecular Engineering, University of Chicago, Chicago, Illinois, USA }
	\author {L. Dos Santos Martins}
	\affiliation{QuTech and Kavli Institute of Nanoscience, Delft University of Technology, Delft, The Netherlands}
	\affiliation{Present address: Sorbonne Université, LIP6, CNRS, Paris, France}
	\author{ A. R.-P. Montblanch}
	\affiliation{QuTech and Kavli Institute of Nanoscience, Delft University of Technology, Delft, The Netherlands}
    \author{H. K. C. Beukers}
    \affiliation{QuTech and Kavli Institute of Nanoscience, Delft University of Technology, Delft, The Netherlands}
    \author{S. Baier}
    \affiliation{QuTech and Kavli Institute of Nanoscience, Delft University of Technology, Delft, The Netherlands}
    \affiliation{Present address: Institut für Experimentalphysik, Universität Innsbruck, Innsbruck, Austria}
    \author{J. Borregaard}
    \affiliation{QuTech and Kavli Institute of Nanoscience, Delft University of Technology, Delft, The Netherlands}
    \author{R. Hanson}
    \affiliation{QuTech and Kavli Institute of Nanoscience, Delft University of Technology, Delft, The Netherlands}
	\date{\today}

\begin{abstract}
	The generation of entanglement between remote matter qubits has developed into a key capability for fundamental investigations as well as for emerging quantum technologies. In the single-photon, protocol entanglement is heralded by generation of qubit-photon entangled states and subsequent detection of a single photon behind a beam splitter. In this work we perform a detailed theoretical and experimental investigation of this protocol and its various sources of infidelity. We develop an extensive theoretical model and subsequently tailor it to our experimental setting, based on nitrogen-vacancy centers in diamond. Experimentally, we verify the model by generating remote states for varying phase and amplitudes of the initial qubit superposition states and varying optical phase difference of the photons arriving at the beam splitter. We show that a static frequency offset between the optical transitions of the qubits leads to an entangled state phase that depends on the photon detection time. We find that the implementation of a Charge-Resonance check on the nitrogen-vacancy center yields transform-limited linewidths. Moreover, we measure the probability of double optical excitation, a significant source of infidelity, as a function of the power of the excitation pulse. Finally, we find that imperfect optical excitation can lead to a detection-arm-dependent entangled state fidelity and rate. The conclusion presented here are not specific to the nitrogen-vacancy centers used to carry out the experiments, and are therefore readily applicable to other qubit platforms.
\end{abstract}
	\maketitle

\section{Introduction}\label{sec:intro}
Entanglement between different nodes will be an essential element of future quantum networks. Entangled states will serve as a key ingredient for many applications, such as secure communication, distributed quantum computation and advanced quantum network protocols \cite{Kimble2008,Wehner2018,Ekert2014,Broadbent2009,Ben-Or2006,Christandl2005,DeBone2020}. Remote entanglement between distant nodes can be generated using different protocols. One of these protocols, the single-photon protocol based on emitted photons encoded in number states \cite{Cabrillo1999, Bose1999}, is especially suited to establish entanglement between distant stationary qubits with high generation rates in the presence of significant photon loss. Since a single photon has to travel only half of the distance between the emitters, the total photon loss is reduced compared to direct photon transmission or to two-photon entangling protocols \cite{Barrett2005}. The single-photon protocol has been implemented on various qubit platforms, such as electron and hole spins in quantum dots, nitrogen-vacancy centers in diamond and atomic ensembles in rare-earth-ion doped crystals \cite{Delteil2016,Stockill2017,Humphreys2018,Lago-Rivera2021}. 

The single-photon protocol works as follows. Two remote qubits are each prepared in the superposition state $\sqrt{\alpha}\ket{0} + \sqrt{1-\alpha}\ket{1}$. State-selective excitation of $\ket{0}$, the so-called \textit{bright} state, creates a qubit-photon entangled state. Interference of the photon states on a balanced beam splitter erases the which-path information. Detection of a single photon projects the two remote qubits into an entangled state $\ket{\Psi} = \frac{1}{\sqrt{2}}(\ket{01} + \ket{10})$, up to a single qubit phase correction. However, in the presence of photon loss we cannot discriminate between the emission of a single photon and the case in which both qubits were in the bright state, $\ket{00}$, and emitted each one photon but one photon was lost. The latter events falsely herald entanglement and reduce the average fidelity. In the high photon-loss regime, given the detection of one photon, the probability that both qubits are in the bright state is given by the initial population in $\ket{0}$, $\alpha$. Hence, the average heralded density matrix is $(1-\alpha) \ket{\Psi}\bra{\Psi} + \alpha\ket{00}\bra{00}$, with a fidelity of $F = 1 - \alpha$ with respect to the maximally entangled state.

Apart from this infidelity intrinsic to the protocol, other sources of error can degrade the heralded state. In this paper we provide a detailed theoretical and experimental study of error sources and characteristics associated with the single-photon protocol. 

The paper is structured as follows. In Section \ref{sec:sce_protocol} we describe the single-photon entanglement protocol step-by-step for a general experimental setting and we develop a model describing the effect of experimental imperfections. We introduce our experimental system, the nitrogen-vacancy (NV) center in diamond, in Section \ref{sec:exp_setup} and we tailor the model to our system in Section \ref{sec:tailored_model}. Afterwards, we discuss the effect of various parameters on the heralded state; the bright state population (Section \ref{sec:bright_state_pop}), the phase of the entangled state (Section \ref{sec:ent_state_phase}), photon indistinguishability (Section \ref{sec:photon_dist}), double optical excitation (Section \ref{sec:double_exc}) and non-excited bright state population (Section \ref{sec:non_pi_pulse}). Several of these experimental parameters are interrelated and in Section \ref{sec:opt} we discuss fidelity and rate optimization of the heralded state.  We present our conclusions in Section \ref{sec:conclusion}.

This work contributes to a better understanding of the effect of general experimental imperfections and its platform-independent insights can be used to improve entanglement generation on various systems, such as other solid state defects and quantum dots~\cite{rose_observation_2018,bhaskar_quantum_2017,nguyen_quantum_2019,trusheim_transform-limited_2020,widmann_coherent_2015,son_developing_2020,dibos_atomic_2018,kindem_control_2020,coste_high-rate_2022}.

\section{The single-photon entanglement protocol}\label{sec:sce_protocol}
\begin{figure}
	\includegraphics{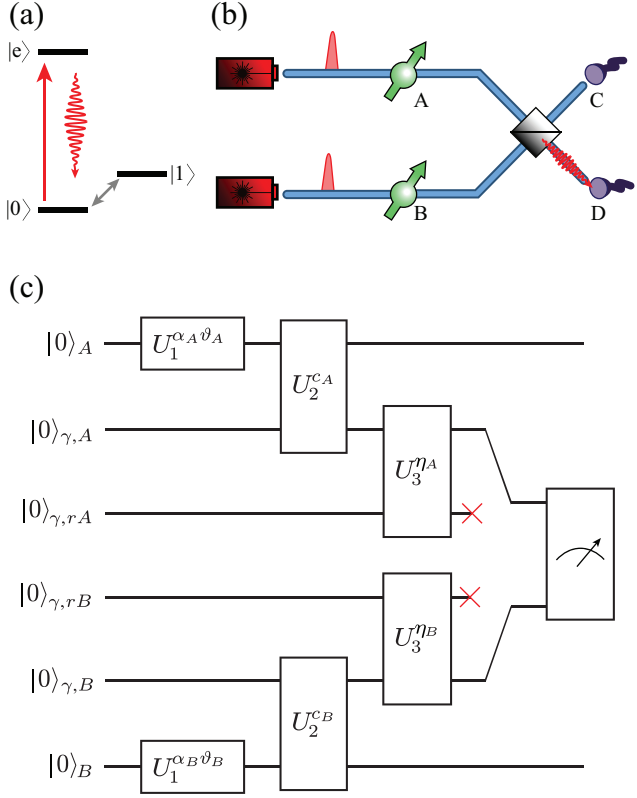}
	\caption{\textbf{Simplified level structure, generalized experimental layout and overview of unitary operations.} \figlet{a} L-scheme level structure for the single-photon entanglement protocol. Two ground states define the qubit subspace $ \ket{0} $ and $ \ket{1} $, and we can drive the transition between them. Furthermore, $\ket{0} $ is optically connected to an excited state $ \ket{e} $. \figlet{b} In a generalized experimental layout, excitation pulses are sent to qubits A and B and the emitted photons are led to a beam splitter. The output ports C and D of the beam splitter are connected to two single photon detectors. \figlet{c} Gate circuit diagram describing the single-photon entanglement protocol. Stationary qubits A and B are initialized into the $\ket{0}_i$ state, where the subscript $i$ denotes qubit A or B. Since there are no photons yet at the start of the protocol, all photon modes are described by the vacuum state, $\ket{0}_{\gamma,i}$. The same holds for the photon modes indicating lost photons, $\ket{0}_{\gamma,r i}$. Throughout the protocol, several unitaries $U$ act on the qubits and the photon states, and finally a state between the stationary qubits is heralded by a joint measurement of the photonic states. In the main text, we describe the different unitaries and the resulting heralded state. }
	\label{fig:levels}
\end{figure}

In this section we provide a step-by-step description of the single-photon protocol and derive the resulting two-qubit state. Figure \ref{fig:levels}a shows an example of the energy levels used by the protocol, in this work we employ a L-scheme for the optical excitation. We would like to emphasize that the protocol can also be executed with a Lambda (or Raman) excitation scheme. In that case, the optical excitation induces the qubit to flip \cite{Cabrillo1999}. Here we develop our model based on the L-scheme, as depicted in Figure \ref{fig:levels}a.  We label two levels $\ket{0}$ and $\ket{1}$ as our qubit subspace, and we can coherently drive the transition between them to create any superposition state. Furthermore, the $\ket{0}$ ground state is connected to an optically excited state $\ket{e}$, allowing for state-selective excitation and qubit-photon entanglement. Figure \ref{fig:levels}b shows a general experimental layout, the two qubits can be individually excited and the emitted photons are led to a beam splitter. The output ports of the beam splitter are connected to two photon detectors. 

\begin{figure}
    \centering
    \includegraphics[width=\linewidth]{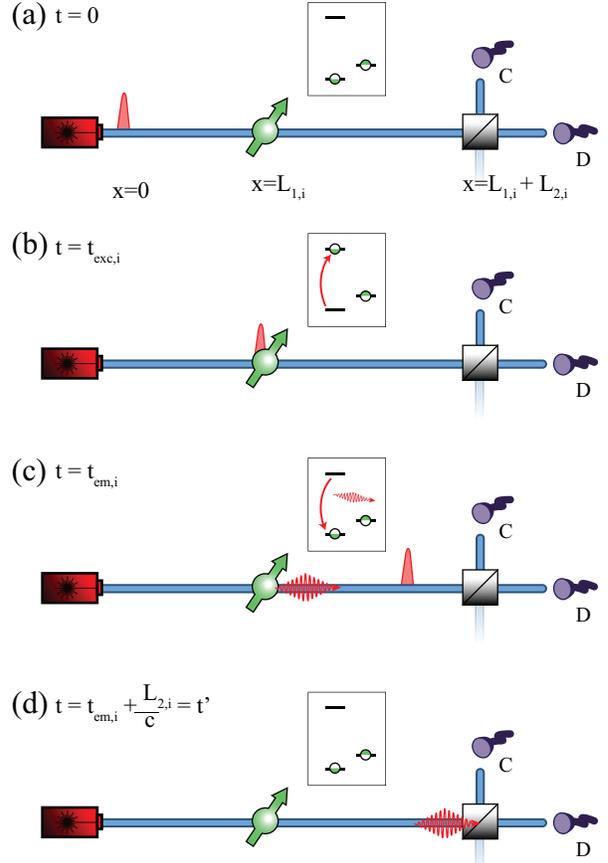}
    \caption{\textbf{Overview of the different time definitions used in the derivation.} \figlet{a} At $t=0$ the excitation pulse is generated at position $x=0$ with phase $\phi_{l,i}$. \figlet{b} At time $t=t_{\text{exc},i}$, the excitation pulse arrives at the emitter and transfers the population from the $\ket{0}$ ground to the excited state $\ket{e}$. \figlet{c} At time $t = t_{\text{em},i}$, the population in the excited state spontaneously decays and a photon is emitted. \figlet{d} Finally, the emitted photon arrives at the detector at position $x = L_{1,i}+L_{2,i}$ and is detected at time $t = t_{\text{em},i} + \frac{L_{2,i}}{c} = t'$}
    \label{fig:time_defs}
\end{figure}

Throughout this section, we describe various unitaries acting on the individual qubits and photons, and we finally discuss a photon detection event heralding the two-qubit state. Figure \ref{fig:levels}c summarizes the protocol in the form of a gate circuit diagram. $\ket{0}_A$ and $\ket{0}_B$ represent the (stationary) qubits at the start of the protocol.  $\ket{0}_{\gamma,A/B}$ describe the photon modes and $\ket{0}_{\gamma,r,A/B}$ the photon modes of any lost photons, both equal to the vacuum state at the start of the protocol as there are no photons yet. Figure \ref{fig:time_defs} provides an overview of the different timing definitions and schematically shows the single-photon entanglement protocol at different points in time.

We work in the rotating frame of both qubits and we start with qubits A and B initialized in the $\ket{0}$ ground state
\begin{equation}
\ket{\Psi_0}_{A} \otimes \ket{\Psi_0}_{B}= \ket{0}_A \otimes \ket{0}_B.
\end{equation}

Next we create a superposition state on each of the qubits using the unitary 
\begin{equation}\label{eq:init_sup_state}
U_1^{\alpha_i, \vartheta_i} : \ket{0} \rightarrow \sqrt{\alpha_i}\ket{0}+\sqrt{1-\alpha_i}e^{-i\vartheta_i}\ket{1}.
\end{equation}
In unitary $ U_1^{\alpha_i, \vartheta_i} $, $\alpha_i$ denotes the population in $\ket{0}$, i.e. the bright state, and $\vartheta_i$ represents the phase of the superposition state. Subscript $i$ denotes either of the two emitters used in the protocol, A or B.

To create the qubit-photon entangled states, we expose the emitter to excitation light. At time $t=0$, a short optical pulse is being created, see Figure \ref{fig:time_defs}a. At time $t_{\text{exc},i}$, qubit $i$ will be state-selectively excited by this optical excitation pulse (Figure \ref{fig:time_defs}b). In this derivation we assume the optical excitation to be instantaneous. Subsequent decay to the ground state will result in the emission of a photon (Figure \ref{fig:time_defs}c) and the photonic state can be written as a superposition of different photon modes $\zeta_i$ with different emission times  $t_{\text{em},i}$

\begin{widetext}
\begin{equation}\label{eq:zeta}
\begin{split}
    \int_{t_{\text{exc},i}}^\infty c_{e,i}\zeta_i(x, t, t_{\text{exc}, i},t_{\text{em},i}) dt_{\text{em},i}  & = \int c_{e,i}\vec{\epsilon}_i \mathcal{E}_i(t_{\text{exc}, i},t_{\text{em},i}) e^{-i(\omega_i(t-t_{\text{em},i}) - k_i(x-L_{1,i}) + \phi_{l,i} + \pi/2 + \omega_i(t_{\text{em},i} - t_{\text{exc},i}))} dt_{\text{em},i} \\
     & = \int c_{e,i}\vec{\epsilon}_i \mathcal{E}_i(t_{\text{exc}, i},t_{\text{em},i}) e^{-i(\omega_i(t-t_{\text{exc},i}) - k_i(x-L_{1,i}) + \phi_{l,i} +  \pi/2)} dt_{\text{em},i}
\end{split}
\end{equation}

In \eref{zeta}, $|c_{e,i}|^2$ is the probability to transfer the population to the excited state $\ket{e}$, $\vec{\epsilon}_i$ denotes the polarization of the emitted photon, $\omega_i$ is the transition frequency $\ket{0} \rightarrow \ket{e}$ with the corresponding wave number $k_i$, $\phi_{l,i} +\pi/2$ is the phase of the excitation laser imprinted on the photon by means of the Rabi drive, $\mathcal{E}_i(t_{\text{exc}, i},t_{\text{em},i})$ is the temporal envelope and $x$ and $t$ as the spatial and time coordinates respectively. For spontaneous decay, this envelope can be modeled as 
\begin{equation}\label{eq:temp_shape}
\mathcal{E}_i(t_{\text{exc}, i},t_{\text{em},i}) = \frac{H(t_{\text{em},i}-t_{\text{exc},i})e^{-(t_{\text{exc},i} - t_{\text{em},i})/2\tau}}{\sqrt{\tau}},
\end{equation}
with $\tau$ the excited state lifetime and $H(t)$ the Heaviside function.

At a later point in the protocol we will describe the detection of a photon at time $t_{\text{em},i} + \frac{L_{2,i}}{c}$ (see Figure \ref{fig:time_defs}d), therefore we can already write the photon state as being emitted in that specific mode
\begin{equation}\label{eq:det_mode}
\begin{split}
c_{e,i}\zeta_i (x,t, t_{\text{exc},i},t_{\text{em},i})= c_{e,i} \vec{\epsilon}_i \mathcal{E}_i(t_{\text{exc}, i},t_{\text{em},i}) e^{-i(\omega_i(t-t_{\text{exc}, i}) - k_i(x-L_{1,i}) + \phi_{l,i} + \pi/2)}\\
\end{split}
\end{equation}
\end{widetext}

 Additionally, in practice, the excitation laser pulse has a finite duration and therefore re-excitation of the qubit is possible in case the first decay happens during the optical pulse (we ignore any higher-order emissions). For this reason we define $t_{\text{exc},2,i}$ and $t_{\text{em},2,i}$ as the times at which the re-excitation and second spontaneous decay occur, with the condition $t_{\text{exc},2,i} > t_{\text{em},i}$.  We write the joint photonic mode as $\zeta_{ii}(t_{\text{exc},i},t_{\text{exc},2,i}) $ and define the double excitation probability as $|c_{ee,i}|^2$ \cite{Das2019,tiurev_fidelity_2021}. Again we assume both photons to be emitted in specific modes. 
 
 
 Together with the probability to remain in the ground state $\ket{0}$, $|c_{0,i}|^2$, the unitary describing the state-selective excitation and emission can be written as
\begin{equation}
U_2^{c_i} : \begin{cases} \begin{split}
\ket{0}\vac \rightarrow \ket{0} \otimes ( & c_{0,i}\vac +  c_{e,i}\zei\hat{a}^\dagger(t_{\text{em},i})\vac + \\
& c_{ee,i}\zeii\hat{a}^\dagger(t_{\text{em},2,i})\hat{a}^\dagger(t_{\text{em},i})\vac), 
\end{split}\\
\ket{1}\vac \rightarrow \ket{1}\vac,
\end{cases}
\end{equation}
where $\hat{a}^\dagger$ is the photon creation operator. Note that we exclude the possibility for off-resonant excitation of any transition related to the $\ket{1}$ ground state. 

The emitted photons are subject to losses, which we assume to be equal for all the photons associated with one of the nodes. We use a beam splitter transformation $U_3^{\eta_i}$ to model photon loss as
\begin{equation}
\begin{split}
U_3^{\eta_i} : \hat{a}^{\dagger}_{i}\ket{0}_\gamma\ket{0}_{\gamma,r} \rightarrow & \sqrt{\eta_i}\hat{a}^{\dagger}_{out,i}\ket{0}_\gamma\ket{0}_{\gamma,r} \\
& + \sqrt{1-\eta_i}\hat{a}^{\dagger}_{r,i}\ket{0}_\gamma\ket{0}_{\gamma,r}.
\end{split}
\end{equation}
$\hat{a}_{r,i}^\dagger$ creates a photon in the loss mode $\ket{0}_{\gamma,r}$ and $\hat{a}_{out,i}^\dagger$ denotes the photon arriving at the beam splitter. We omit the \textit{out} subscript for brevity. In $U_3^{\eta_i}$, $|\eta_i|^2$ represents the photon detection probability. The (separable) state in front of the central beam splitter will then be given by
\begin{equation}
\begin{split}
	\ket{\Psi_3}_{AB} =  & U_3^{\eta_A} U_2^{c_A} U_1^{\alpha_A, \vartheta_A}\ket{0}_A\ket{0}_{\gamma,A}\ket{0}_{\gamma,rA} \\
	& \otimes U_3^{\eta_B} U_2^{c_B}  U_1^{\alpha_B, \vartheta_B}\ket{0}_B\ket{0}_{\gamma,B}\ket{0}_{\gamma,rB}.
\end{split}
\end{equation}

Let us now turn to the detection.  We assume non-number resolving single photon detectors. Due to photon loss, double excitation and the presence of noise counts, different detection patterns will look identical. As a consequence, we will herald an average state, averaged over the different possible detection patterns. However, we can reject any experimental repetition in which two photons are detected in different detectors. These events suggest either double excitation or both qubits being in the bright state, deteriorating the average state compared to the maximally entangled. Furthermore, assuming the presence of modest photon loss, we consider at most two photons arriving at the beam splitter. We treat all the detection patterns separately and later on combine the result to obtain the average heralded density matrix. We assume the loss after the central beam splitter to be symmetric, for this reason we treat any loss after the beam splitter as an additional contribution to the loss handled by $U_3^{\eta_i}$. 

\textit{Single photon - } The first detection pattern we consider is the case where one photon is emitted and one photon is detected. For a detection at time $t'= t_{\text{em},A} + \frac{L_{2,A}}{c} = t_{\text{em},B} + \frac{L_{2,B}}{c}  $ in port C of the beam splitter, we model the detection by projecting the state onto $\bra{0}_\gamma \hat{a}_{C}(t') $, where $\hat{a}_{C}(t')$ is the photon annihilation operator acting at time $t'$. The effect of the beam splitter is mapping the state to a superposition of a photon originating from A or B
\begin{equation}
\begin{split}
\bra{0}_\gamma \hat{a}_{C}(t')  = \frac{1}{\sqrt{2}}\bra{0}_\gamma\left(\hat{a}_{A}(t') + \hat{a}_{B}(t')\right).
\end{split}
\end{equation}
Similarly we can define the projector for detecting a photon in port D as 
\begin{equation}
\begin{split}
\bra{0}_\gamma \hat{a}_{D}(t') = \frac{1}{\sqrt{2}}\bra{0}_\gamma\left(\hat{a}_{A}(t') - \hat{a}_{B}(t')\right).
\end{split}
\end{equation}

For detecting a photon in port C, the corresponding (unnormalized) density matrix is given by
\begin{equation}
	\rho_1 = \bra{0}_\gamma \hat{a}_{C}(t') \ket{\Psi_3}\bra{\Psi_3}_{AB}  \hat{a}_{C}^\dagger(t') \vac.
\end{equation}

\textit{Two photons - } Secondly, we deal with the case where two photons are emitted, either both by the same node or by the two nodes separately, and both photons arrive at the beam splitter. As mentioned above, two-photon events will be rejected as they suggest both qubits to be in the bright state or double optical excitation. Therefore, we reject events where two photons are being detected in different ports of the beam splitter. When two photons leave the beam splitter via the same port, we cannot discriminate a two-photon event from a single photon event due to the non-number resolving detectors and the first photon will herald a state. In this case, the first photon will be detected in port C at time $t'$ and the second photon will be similarly absorbed at (unknown) time $t'' $, also in port C. The projector is then given by
\begin{equation}
\begin{split}
\bra{0}_\gamma \hat{a}_{C}(t')  \hat{a}_{C}(t'')  = \frac{1}{2}\bra{0}_\gamma(\hat{a}_{A}(t') & + \hat{a}_{B}(t')) (\hat{a}_{A}(t'') + \hat{a}_{B}(t'')),
\end{split}
\end{equation}
and we obtain 
\begin{equation}
\rho_2 = \bra{0}_\gamma \hat{a}_{C}(t')\hat{a}_{C}(t'') \ket{\Psi_3}\bra{\Psi_3}_{AB}  \hat{a}_{C}^\dagger(t'')\hat{a}_{C}^\dagger(t') \vac,
\end{equation}
again an unnormalized density matrix.

\begin{widetext}
\textit{At least one lost photon - } When at least one photon is lost, detection of any remaining photon will falsely herald entanglement. This cannot lead to coherence between the qubit states and therefore we can project on each detection pattern separately and sum over the resulting density matrices
\begin{equation}
\begin{split}
\rho_{\text{incoherent}} = &   \bra{0}_{\gamma} \hat{a}_{C}(t') \hat{a}_{rA}(t_r) \ket{\Psi_3}\bra{\Psi_3}_{AB} \hat{a}_{rA}^\dagger(t_r)\hat{a}_{C}^\dagger(t') \vac \\
& +  \bra{0}_{\gamma} \hat{a}_{C}(t') \hat{a}_{C}(t'') \hat{a}_{rA}(t_r) \ket{\Psi_3}\bra{\Psi_3}_{AB} \hat{a}_{rA}^\dagger(t_r)\hat{a}_{C}^\dagger(t'') \hat{a}_{C}^\dagger(t') \vac \\
& + ... \\
\end{split}
\end{equation}
Since the detection time of the lost photon $t_r$ and therefore its emission time $t_{\text{em,i}}$ is unknown, we can integrate over all possible emission times;  $|\int c_{e,i}\zeta_{i}(t_{\text{em,i}}) d t_{\text{em,i}}|^2 = |c_{e,i}|^2$. 
\end{widetext}

\textit{Noise photon - } A detector dark count or stray light will additionally lead to falsely heralded entangled states. We assume the contribution of such noise counts small compared to actual signal photons and we ignore the (small) probability of both a signal photon and noise photon arriving at the detector. We distinguish two scenarios: no photon is emitted, or none of the emitted photons arrived at the beam splitter. The first scenario leads to a separable state, but the single qubit coherence is not lost, while the latter projects the corresponding qubit state and its coherence is lost. We can deal with these two scenarios separately. Regarding the first scenario, by projecting on the vacuum state $\vac$, we get the density matrix
\begin{equation}
	\rho_{0} = \ket{\Psi_{0}} \bra{\Psi_{0}}_A \otimes \ket{\Psi_{0}} \bra{\Psi_{0}}_B,
\end{equation}
with 
\begin{equation}
\begin{split}
\ket{\Psi_{0}}_i = (\sqrt{\alpha_i} c_{0,i}\ket{0} + e^{-i\vartheta_i}\sqrt{1-\alpha_i} \ket{1}).
\end{split}
\end{equation}
Alternatively, when none of the emitted photons arrive at the beam splitter, we can trace over all the lost photons to obtain
\begin{equation}
\begin{split}
\rho_{\text{lost}} = & \bra{0}_{\gamma} \hat{a}_{rA}(t_r) \ket{\Psi_3}\bra{\Psi_3}_{AB} \hat{a}_{rA}^\dagger(t_r) \vac \\
 & + \bra{0}_{\gamma} \hat{a}_{rA}(t_r) \hat{a}_{rA}(\tilde{t_r}) \ket{\Psi_3}\bra{\Psi_3}_{AB} \hat{a}_{rA}^\dagger(\tilde{t_r}) \hat{a}_{rA}^\dagger(t_r) \\
 & + ... \\
\end{split}
\end{equation}
We add the two noise contributions and multiply by the probability for a noise count being detected, $p_d$ and construct 
\begin{equation}
\rho_{\text{noise}} = p_d (\rho_{0} + \rho_{\text{lost}}).
\end{equation}

Having discussed all the detection patterns we can finally combine all the contributions to extract the probability to get a detection event in port C
\begin{equation}\label{eq:pclick}
\begin{split}
p_{\text{click,C}} = & \Tr(\rho_{1}) +\Tr(\rho_{2}) +\Tr(\rho_{\text{incoherent}})+\Tr(\rho_{\text{noise}}).\\
\end{split}
\end{equation}
We add all the density matrices and normalize using the detection probability $ p_{\text{click,C}} $ and we obtain an expression for the average density matrix $\rho_C$ heralded by a detection in port C
\begin{equation}\label{eq:final_rho}
\begin{split}
\rho_C = &\frac{1}{p_{\text{click,C}}} (\rho_{1} + \rho_{2} +\rho_{\text{incoherent}} + \rho_{\text{noise}}).\\
\end{split}
\end{equation}
In Appendix \ref{app:full_state}, we provide the full description of the density matrix. 

In the introduction we discussed the expected fidelity in the high loss regime and here we check that our model matches the intuitive result. We use a  shorthand notation for the detected photon modes defined in \eref{det_mode} as $\zeta_{i}(t')$, with $t'$ ($t''$) the detection time of the first (second) photon. In the limit of high photon loss ($\eta \ll 1$), equal experimental settings ($\eta_A = \eta_B \equiv \eta$ and $\alpha_A = \alpha_B \equiv \alpha$), perfect optical excitation pulses $|c_{e,i}|^2 = 1$, perfectly indistinguishable photons with equal optical phase upon arrival at the beam splitter $\zeta_A(t') = \zeta_B(t') \equiv \zeta(t')$ and absence of noise counts $p_d$, we can simplify the results to
\begin{equation}\label{eq:pclick_high_loss}
\begin{split}
p_{\text{click},\eta \ll 1,C}  = & \Tr(\rho_{1}) +\Tr(\rho_{\text{incoherent}})\\
= &\alpha(1-\alpha)\eta|\zeta(t')|^2 + \alpha^2\eta|\zeta(t')|^2\\
= & \alpha\eta|\zeta(t')|^2,\\
\end{split}
\end{equation}
where $\eta|\zeta(t')|^2$ can be interpreted as the probability to detect a photon in the detection window. The density matrix can be written as 
\begin{align}
\begin{split}
\rho_{\eta \ll 1,C} =& \frac{1}{p_{\text{click},\eta \ll 1,C}} (\rho_{1} + \rho_{\text{incoherent}}), \\
= & \frac{\alpha(1-\alpha)\eta|\zeta(t')|^2}{p_{\text{click},\eta \ll 1,C}} \ket{\Psi}\bra{\Psi} + \frac{\alpha^2\eta|\zeta(t')|^2}{p_{\text{click},\eta \ll 1,C}}  \ket{00}\bra{00},\\
= & (1-\alpha)\ket{\Psi^+}\bra{\Psi^+} + \alpha\ket{00}\bra{00},\\
\end{split}
\end{align}
with $\ket{\Psi^+} = \frac{1}{\sqrt{2}}(\ket{01} + \ket{10})$, the maximally entangled Bell state, matching our intuitive prediction presented in Section \ref{sec:intro}. Similarly, a photon detection in port D will give the same result, albeit with the entangled state $\ket{\Psi^-} = \frac{1}{\sqrt{2}}(\ket{01} - \ket{10})$.

\section{Experimental setup: NV centers}\label{sec:exp_setup}
In this work we use nitrogen-vacancy centers (NV) in bulk diamond as our qubit system. This defect consists of a substitutional nitrogen-atom with the adjacent lattice site left vacant. In the negative charge state an additional electron from the environment is trapped and a spin-1 system is formed. We use two ground states, $\ket{0} = \ket{m_s = 0}$ and $\ket{1} = \ket{m_s = -1}$ (or $\ket{m_s = +1}$), as our qubit subspace. The $\ket{m_s = 0}$ state is connected to an optically excited state $\ket{e} = \ket{E_x}$ \footnote{In one of the experimental setups, node C, we use $\ket{E_y}$ as the excited state.}, and the transition can be selectively addressed at cryogenic temperatures \cite{Robledo2011a, Robledo2011}. Spontaneous decay from the excited state $\ket{e}$ to $\ket{0}$ happens resonantly $\approx 3\%$ of the time, into the so-called zero-phonon line (ZPL) \cite{Riedel2017}. In the remaining $\approx 97\%$, the decay happens off-resonantly into the phonon-side band (PSB); in this case the emitted photon is accompanied with the emission of a phonon. Tracing over the (undetected) phonon state erases the spin-photon coherence, prohibiting PSB photons to be used for entanglement generation. In the optical setup we separate the ZPL from the PSB photons using a dichroic mirror \cite{Pompili2021}.

In this work, we use the same experimental setups as used in the quantum network of References \cite{Pompili2021} and \cite{Hermans2021}. We label the nodes Alice (A), Bob (B) and Charlie (C), which are connected in such a way that we can generate entanglement on two links; Alice-Bob (AB) and Bob-Charlie (BC). The exact connections are given in \cite{Pompili2021}. By applying a DC voltage via on-chip electrodes, we tune the optical transitions of nodes A and C to match the frequency of node B using the DC Stark effect \cite{Tamarat2006}. For each link the nodes share the excitation laser and the short excitation pulses are generated using an electro-optic modulator (EOM, Jenoptik) driven by an arbitrary waveform generator (AWG, Z\"urich Instruments). To provide additional extinction of the pulse, we make use of an acousto-optic modulator (AOM, Gooch\&Housego). We use microwave (MW) pulses to drive the transition between the qubit states and with I- and Q-modulation we can generate any superposition. We actively stabilize the optical phases acquired by the excitation pulses and photons using a combination of heterodyne and homodyne phase detection methods and feedback \cite{Pompili2021}. 

To perform a readout on one of the individual qubit, the qubit is rotated to the required basis using a MW pulse and the NV center is read out by exposure to light resonant with the $\ket{0} \rightarrow \ket{e}$ transition. Detection of a photon in the PSB detection path marks a $\ket{0}$ state readout, while absence of photons corresponds to the $\ket{1}$ state.

The optical transition frequencies of the NV center are sensitive to (laser-induced) changes in the charge-environment~\cite{robledo_control_2010,Doherty2011,orphal-kobin_optically_2022,mccullian_quantifying_2022}. To mitigate this effect we perform a Charge-Resonance (CR) check prior to every experimental run \cite{Robledo2011}. During a CR check, we turn on the control lasers to ensure the emitter is on resonance with the control lasers and in the correct charge state. Only when a number of PSB photons above a pre-set threshold is detected, an experimental repetition is started. 

\section{Tailoring the model for NV centers}\label{sec:tailored_model}
In Section \ref{sec:sce_protocol} we have considered the single-photon protocol in a general way. Considering our experimental implementation using NV centers in bulk diamond, we can make several approximations to simplify the results. Due to the small fraction of resonantly emitted photons ($\approx 3\%$) and limited detection efficiency ($<15\%$), we can assume $\eta \ll 1$. By using high-power laser pulses we can assume all population to be transferred to the optically excited state, $c_{0,i} = 0$. We set a detection window which starts after the arrival time of the optical pulse to mitigate counts due to imperfectly rejected laser light reaching the detectors. As a consequence, during a double excitation event the first photon will never be detected and we set $\eta_i(1-\eta_i)|\zeta_{ii}(t', t_r)|^2 = 0$. In this way, we can define the parameter $p_{de}$ as the probability that a second photon is emitted given a photon detection: $|c_{ee,i}\zeta_{ii}(t_r,t')|^2 = p_{de}$. 

With these assumptions we can simplify \eref{pclick} and (\ref{eq:final_rho}), and we obtain for the probability to detect a photon in port C
\begin{equation}
\begin{split}
	p_{\text{click,NV,C}} = & \Tr(\rho_{1}) + \Tr(\rho_{\text{incoherent}}) + \Tr(\rho_{\text{noise}}).\\
\end{split}
\end{equation}
The heralded density matrix is then given by
\begin{equation}
\begin{split}
\rho_{\text{NV,C}} =& \frac{1}{p_{\text{click,NV,C}}} (\rho_{1} + \rho_{\text{incoherent}} + \rho_{\text{noise}}) \\
= &\frac{1}{p_{\text{click,NV,C}}} \begin{pmatrix}
a_{00} & 0 & 0 & 0 \\
0 & a_{11}  & a_{12} & 0 \\
0 & a_{12}^\ast & a_{22} & 0 \\
0 & 0 & 0 & a_{33} \\
\end{pmatrix}
\end{split}
\end{equation}
with the elements
\begin{align}
\begin{split}
a_{00} = &  \alpha_A\alpha_B (\frac{1}{2}\eta_A |c_{e,A}\zea(t')|^2  + \frac{1}{2}\eta_B  |c_{e,B}\zeb(t')|^2 \\
& + \frac{1}{2}\eta_A p_{de}  + \frac{1}{2}\eta_B p_{de} + p_d ),\\
\end{split}
\\
\begin{split}
a_{11} = \alpha_A(1-\alpha_B)(\frac{1}{2} \eta_A |c_{e,A}\zea(t')|^2 +  \frac{1}{2}\eta_A p_{de} + p_d ),\\
\end{split}
\\
\begin{split}
a_{22} = (1-\alpha_A)\alpha_B( \frac{1}{2} \eta_B |c_{e,B}\zeb(t')|^2 + \frac{1}{2}\eta_B p_{de} + p_d),\\
\end{split}
\\
\begin{split}
a_{12} =  \frac{1}{2}M e^{-i\phi},\\
\end{split}
\\
\begin{split}
a_{33} =  p_d (1-\alpha_A)(1-\alpha_B ).\\
\end{split}
\end{align}
In the expression of $a_{12}$, $M$ is the magnitude of the coherence term
\begin{equation}\label{eq:off_diag_magnitude}
	\begin{split}
M =& \sqrt{\alpha_A(1-\alpha_A)\alpha_B(1-\alpha_B)\eta_A\eta_B} \\
& \times (\vec{\epsilon_A}\cdot\vec{\epsilon_B}) c_{e,A}c_{e,B}\mathcal{E}_A(t_{\text{exc},A}, t_{\text{em},A})\mathcal{E}_B(t_{\text{exc},B}, t_{\text{em},B}), \\
	\end{split}
\end{equation} 
and $\phi$ represents the phase of the entangled state
\begin{equation}\label{eq:ent_state_phase}
\begin{split}
\phi = & \vartheta_B-\vartheta_A - \omega_A (t_{\text{em},A}-t_{\text{exc},A}) \\
& + \omega_B (t_{\text{em},B}-t_{\text{exc},B})  -\phi_{l,A} + \phi_{l,B}. \\
\end{split}
\end{equation}

\begin{figure}[t]
	\centering
	\includegraphics[width=\linewidth]{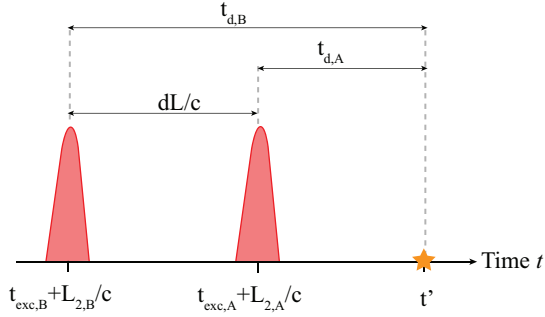}
	\caption{\textbf{Photon detection time.} The photon detection happens at time $t = t'$. We introduce an extra variable, $ t_{d,i} $ with $i = (A,B)$, which is defined as the time spent in the excited state $t_{\text{em,i}} - t_{\text{exc,i}}$. $t_{d,i}$ can also be viewed as the photon detection time with respect to the arrival times of the optical pulses, which occur at $t_{\text{exc,i}} + L_{2,i}/c$ (with $c$ the speed of light constant). The arrival time difference between the optical pulses depends on the optical path length difference $dL = L_{1,A} + L_{2,A} - L_{1,B} - L_{2,B}$, and is given by $dL/c$. }
	\label{fig:arrival_times}
\end{figure}

In our experimental implementation, the emitters share the excitation laser and the optical paths are much smaller than the coherence length of the laser. Thus, we can rewrite \eref{ent_state_phase}. The phase of the laser is $\phi_{l,A} = \phi_{l,B} \equiv \phi_l$ and the frequency of the laser $\omega_{l,A} = \omega_{l,B} \equiv \omega_l$. We introduce a detuning $\Delta_i$ between the laser and the optical transitions, $\omega_l - \omega_i = \Delta_i$ of the emitters ($i=A,B$). Even though the emitters share the same excitation pulse, the path lengths of the excitation pulse and single photons do not have to be same and their difference is given by $ L_{1,A} + L_{2,A} - L_{1,B} - L_{2,B} = dL $. Hence, for a photon detection event at time $t'$, the time spent in the excited state can be different for the two emitters, $t_{\text{em},A}-t_{\text{exc},A} \neq t_{\text{em},B}-t_{\text{exc},B}$. We introduce the variable $t_{d,i}$ as the time spent in the excited state, $t_{\text{em},i}-t_{\text{exc},i}$, see Figure \ref{fig:arrival_times}. Note that $t_{d,i}$ can also be viewed as the detection time with respect to the arrival time of the excitation pulse at the beam splitter. Using these definitions we can write $\phi$ as 
\begin{equation}
\begin{split}
\phi = & \vartheta_B-\vartheta_A - \omega_A t_{d,A} + \omega_B t_{d,B} \\
=& \vartheta_B-\vartheta_A -\omega_l(t_{d,A}-t_{d,B}) + \Delta_A t_{d,A} - \Delta_B t_{d,B} \\
= & \vartheta_B-\vartheta_A + \omega_l\frac{dL}{c} + \Delta_A t_{d,A} - \Delta_B t_{d,B} 
\end{split}
\end{equation}
where $\omega_l\frac{dL}{c}$ is the optical phase difference between the paths, $c$ denotes the speed of light. As mentioned in Section \ref{sec:exp_setup}, we stabilize this phase difference to a setpoint  $\delta\varphi$. We can rewrite $t_{d,B}$ as a function of $t_{d,A}$; $t_{d,B} = t_{d,A} + \frac{dL}{c}$ and obtain
\begin{equation}\label{eq:ent_phase_temp}
\begin{split}
\phi =& \vartheta_B-\vartheta_A+\delta\varphi + \Delta_A t_{d,A} - \Delta_B t_{d,A} - \Delta_B \frac{dL}{c}).\\
\end{split}
\end{equation}
The last term in \eref{ent_phase_temp} can be ignored when either the arrival time difference $\frac{dL}{c}$ or the detuning $\Delta$ of one of the nodes is small. Assuming $\frac{dL}{c} \ll t_{d,A}$, we can write the entangled state phase as
\begin{equation}\label{eq:ent_phase_shared}
\phi =  \vartheta_B-\vartheta_A+\delta\varphi+ \Delta_A t_{d} - \Delta_B t_{d}.
\end{equation}
where $t_{d}$ is the photon detection time with respect to the (nearly equal) arrival time of the optical pulses to the detectors.

In the remainder of the paper, we use this model to simulate the fidelity with respect to the maximally entangled state and the success probability. The simulations include a Monte Carlo simulation to incorporate the exponential probability distribution of the photon detection times, as well as Gaussian probability distributions of the noise of the phase stabilization setpoint and frequency detunings ~\cite{software_and_data}.

\section{Bright state population}\label{sec:bright_state_pop}

\begin{figure}
	\includegraphics{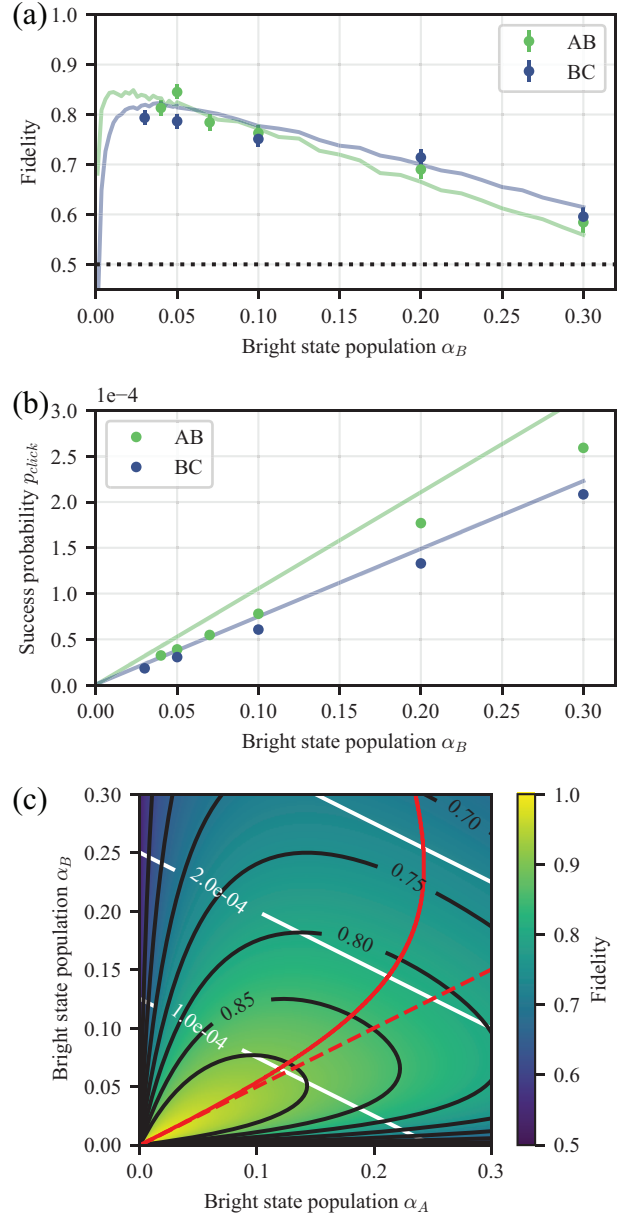}
	\caption{\textbf{Effect of bright state population.} Fidelity with respect to the maximally entangled state \figlet{a} and probability to herald a state \figlet{b} as a function of the bright state populations $ \alpha $. The data (circles) is measured on two links of the network of Reference \cite{Pompili2021}, AB and BC. The x-axis represents the bright state population of setup B in both cases. The bright state population of node A is scaled to be $\alpha_A = \frac{\eta_B}{\eta_A}\alpha_B$, while $\alpha_C = \alpha_B$. The solid lines are given by our model, see Table \ref{tab:exp_params} for the parameters. \figlet{c} In absence of errors other than the protocol error, we calculate the entangled state fidelity for various settings of $ (\alpha_A,\alpha_B) $ for the case $ \eta_B= 2\eta_A$ with $\eta_A =$ 4e-4. The black and white solid lines are isolines for the entangled state fidelity and success probability respectively. The red solid line indicates the optical settings to obtain the highest fidelity with respect to the maximally entangled state for a fixed success probability. The red dashed line represents the $\eta_A\alpha_A = \eta_B\alpha_B$ scenario. }
	\label{fig:brightness_imbalance}
\end{figure}
In this section we study the effect of the bright state population $\alpha$. First, we will vary $\alpha$ and discuss the effect on the fidelity of the heralded state in the presence of noise. Second, we will discuss optimal settings for the individual $\alpha_i$ when the detection efficiencies of the nodes are not the same. 

In a practical experimental setting, entanglement might be falsely heralded by noise counts. The noise counts can originate from different sources, such as dark counts of the detector, excitation light leaking into the detectors or stray light. The effect on the average heralded state depends on the ratio between noise and signal photons. In Figure \ref{fig:brightness_imbalance}a we plot the measured and simulated fidelity with the maximally entangled state for various settings of $\alpha$. For high values of $\alpha$ we observe the linear scaling of the fidelity with $\alpha$, as suggested by the model for the high-photon loss regime $\eta \ll 1$. For low values of $\alpha$, the fidelity deviates from the linear behavior and for sufficiently low $\alpha$ we observe a sharp drop-off, indicating a significant contribution of falsely heralding noise counts. We note that off-resonant excitation of undesired transitions or errors in the preparation of the superposition state (unitary $U_1^{\alpha, \vartheta}$) can also lead to a sharp drop-off in fidelity for low values of $\alpha$. In Figure \ref{fig:brightness_imbalance}b we plot the success probability $p_{\text{click}}$ as a function of $\alpha$, displaying the  expected linear behavior (see \eref{pclick_high_loss}). 

We now turn to the individual detection efficiency $ \eta$. Generally, the detection efficiency or loss parameter, $\eta$, is not the same for the two nodes due to differences in the individual experimental setups or unequal fiber loss in the paths from the nodes to the beam splitter. For this reason we would like to find optimal settings for $ \alpha_A $ and $ \alpha_B $ to establish remote entanglement with the highest fidelity for a fixed success probability $ p_{click} $. In other words, we want to optimize $ F(\alpha_A, \alpha_B) $ subject to $ \eta_A\alpha_A +\eta_B\alpha_B = p_{click} $ in the high-loss regime. For simplicity, we assume no errors other than the protocol error (i.e. no noise counts, no double excitation, perfectly indistinguishable photons), set an entangled state phase of $\phi=0$ and integrate over all possible detection times such that $\int|\zeta_i(t')|^2 dt' = 1$. To compute the fidelity we calculate the overlap with the maximally entangled state
\begin{align}
	\begin{split}
	F =& \bra{\Psi}\rho_C \ket{\Psi}, \\
	=& \frac{1}{4p_{click,C}}(a_{11} + a_{22} + 2|a_{12}|),\\
	=& \frac{1}{4p_{click,C}}(\alpha_A\eta_A(1-\alpha_B) +\alpha_B\eta_B(1-\alpha_A) \\
	& + 2\sqrt{\alpha_A\alpha_B(1-\alpha_A)(1-\alpha_B)} ),\\
	\end{split}
\end{align}
and the success probability per attempt
\begin{equation}
\begin{split}
	p_{click} = & a_{00} + a_{11} + a_{22},\\
	= & \frac{1}{2}(\eta_A\alpha_A + \eta_B\alpha_B).\\
\end{split}
\end{equation}
To find the optimal settings, we use a Langrangian formalism. The optimal settings are met when the gradients of the fidelity and success probability are parallel, i.e. when the fidelity is maximal for a fixed success probability. We can write this condition as
\begin{equation}\label{eq:opt_condition}
	\begin{split}
	(\nabla p_{click})_\perp \cdot \nabla F &= 0\\
	-\eta_B\frac{\partial F }{\partial \alpha_A} + \eta_A\frac{\partial F }{\partial \alpha_B} &=0.\\
	\end{split}
\end{equation}
We solve \eref{opt_condition} numerically for the case $ \eta_B = 2\eta_A$, with $\eta_A =$ 4e-4. In \figref{brightness_imbalance}c we plot the entangled state fidelity for different values of $ \alpha_A $ and $ \alpha_B $. The optimal settings are represented by the red solid line. For high-fidelity states, the optimal settings are close to $ \alpha_A \eta_A = \alpha_B \eta_B $ (red dashed line). We can interpret this result as balancing the probability of the photon to originate from either setup. For low-fidelity states, the optimal settings differ from balancing the detection probabilities. This can be explained with a simple example; by setting $ \alpha_A = 0 $ and $ \alpha_B \neq 0 $, the detected photon will always originate from setup B and thus we will measure perfect classical anti-correlations. However any quantum correlations are completely washed out and we obtain a fidelity $ F = 0.5 $, irrespective of the value of $ \alpha_B $.  On the contrary, if we set both $ \alpha_A, \alpha_B \neq 0 $ we do get quantum correlations but the protocol error can now push the fidelity below 0.5. Hence, for low-fidelity states the optimal settings for $ \alpha_A $ and $ \alpha_B $ optimize classical anti-correlation at the expense of quantum correlations. 


\section{Phase of the entangled state}\label{sec:ent_state_phase}

\begin{figure*}
	\includegraphics{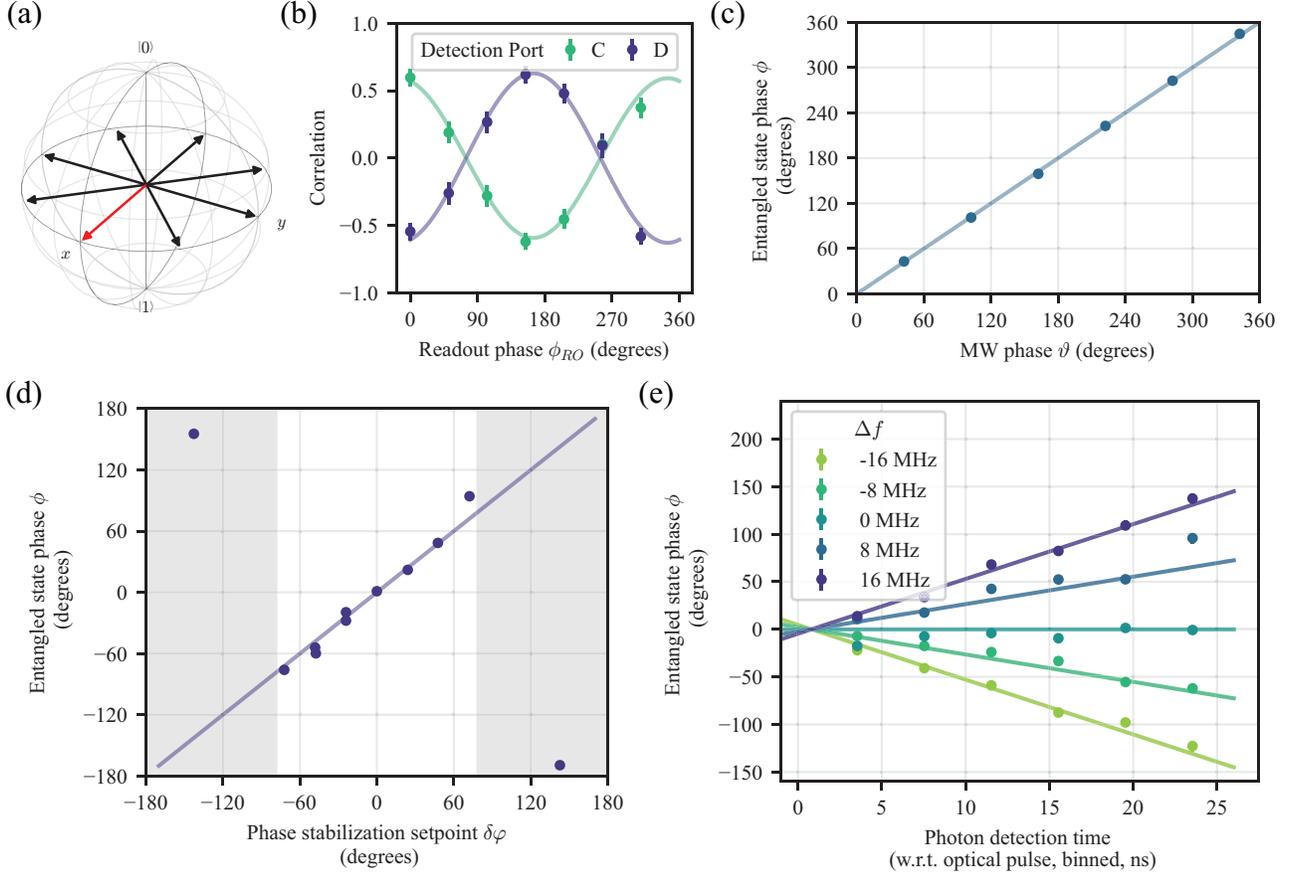}
	\caption{\textbf{Entangled state phase.} \figlet{a} We determine the entangled state phase by measuring along the $ +X $ axis on setup B (red arrow), while sweeping the measurement axis of setup C over the equator plane of the Bloch sphere (black arrows). In this way we obtain oscillating correlations between the measurement outcomes, as shown in \figlet{b}, for states heralded by detector C (turquoise circles) and D (purple circles). We jointly fit the data for the two detectors to extract $ \phi $, the phase offset with respect to a cosine.  \figlet{c} The entangled state phase $ \phi $ as a function of the phase of the microwave (MW) pulse we use to create the initial superposition state (see \eref{init_sup_state}). Here we fix $ \vartheta_C = 0 $ and sweep $ \vartheta_B $. Due to a small difference of the stabilized path compared to the path of the excitation pulses and the single photons, there is a nonzero phase offset \cite{Pompili2021}. We fit the measured phase values with a line with slope 1 and subtract the fitted offset. The solid line shows the fit with the subtracted offset. \figlet{d} The entangled state phase as a function of the optical phase stabilization setpoint. Again we account for the phase offset due to path difference and show the fit with the subtracted offset (solid line). \figlet{e} The entangled state phase as function of the detection time of the photon for different frequency offsets $\Delta f$ between the setups. The photon detection times are binned in bins of 4 ns, and the x value represents the middle of the bin. The time scale is with respect to the highest intensity point of the optical pulse. The solid lines are a joint fit to the data with the zero point crossing $x_0$ as the only free parameter. The fit gives $x0 = (0.8 \pm 0.3)$ ns. The error bars of the data plotted in (C), (D) and (E) are smaller than the symbol size.}
	\label{fig:ent_state_phase}
\end{figure*}

In Section \ref{sec:tailored_model} we discussed the phase of the entangled state $\phi$ and derived the expression for the phase in the case where the excitation laser is shared between the emitters, \eref{ent_phase_shared}. Here we experimentally verify the effect of the different parameters on the entangled state phase, using nodes B and C.

We measure the phase of the entangled state by sweeping the readout basis of node C over the XY-plane of the Bloch sphere (black arrows in Figure \ref{fig:ent_state_phase}a) while we fix the readout basis of node B to be along the +X axis (red arrow). In Figure \ref{fig:ent_state_phase}b we plot the correlations of the measured readout outcomes as a function of the readout basis of node C for states heralded by detecting a photon in port C (turquoise circles) or port D of the beam splitter (purple circles). To extract the phase we jointly fit the two curves and extract $\phi$, the phase offset with respect to a cosine.

 First we vary the phase of the microwave pulse that creates the superposition state on setup B, i.e.~ we change $ \vartheta_B $, the phase of the initial superposition state. We fit the data with a line with slope 1 and subtract the fitted offset. The data with the subtracted offset and the fit are are plotted in \figref{ent_state_phase}c and we observe the expected linear dependence.

 The next parameter we vary is the setpoint for the optical phase stabilization to change $\delta\varphi$.  In \figref{ent_state_phase}d we plot the measured entangled state phase $ \phi $ (again with a fitted offset subtracted) for different setpoints of the stabilization (circles) together with the fit with a fixed slope 1 (solid line). We use the phase stabilization architecture of Reference~\cite{Pompili2021}. In this architecture, the optical phase difference $\delta \varphi$ is governed by a linear combination of the setpoints of the three individual interferometers. This enables us to stabilize to values of $\delta \varphi$ further away from 0, but ultimately the non-linear sinusoidal phase signals hinder effective stabilization. In the shaded regime of \figref{ent_state_phase}d, the slopes of the individual phase signals are below 0.9 and we assume this regime to be insufficiently linear for effective stabilization.  
 
The third parameter we modify is the frequency difference between the emitters of each setups. We change the frequency of the excitation laser and shift the resonance condition of one of the emitters to the new frequency using the DC Stark effect (we apply a DC voltage via on-chip electrodes), while leaving the other emitter at its original emission frequency (thus introducing a detuning to the excitation laser). As derived in \eref{ent_phase_shared}, a frequency difference results in a shift of the entangled state phase depending on the detection time of the photon. This dependence of phase on the photon detection time has also been observed in~\cite{zhao_entangling_2014,vittorini_entanglement_2014}. In \figref{ent_state_phase}e we plot the measured entangled state phase as a function of the detection time of the photon (in bins of 4 ns) for various frequency offsets between the emitters. The time axis is with respect to the highest intensity point of the excitation pulse. We perform a combined fit of the data with fixed slopes given by our model. In a quantum jump picture, the fitted crossing of the lines can be interpreted as the average point in time where the excitation from $\ket{0}$ to $\ket{e}$ occurs, and thus the starting point of the time spent in the excited state. For large frequency detunings (comparable to the inverse of the pulse width), the effective averaging over different excitation times could also lead to a decrease of the average fidelity, however we expect this contribution to be small for our pulse width (2 ns). The observed dependence of the entangled state phase on the photon detection time is consistent with our model.

\section{Photon distinguishability}\label{sec:photon_dist}

\begin{figure*}
	\includegraphics[]{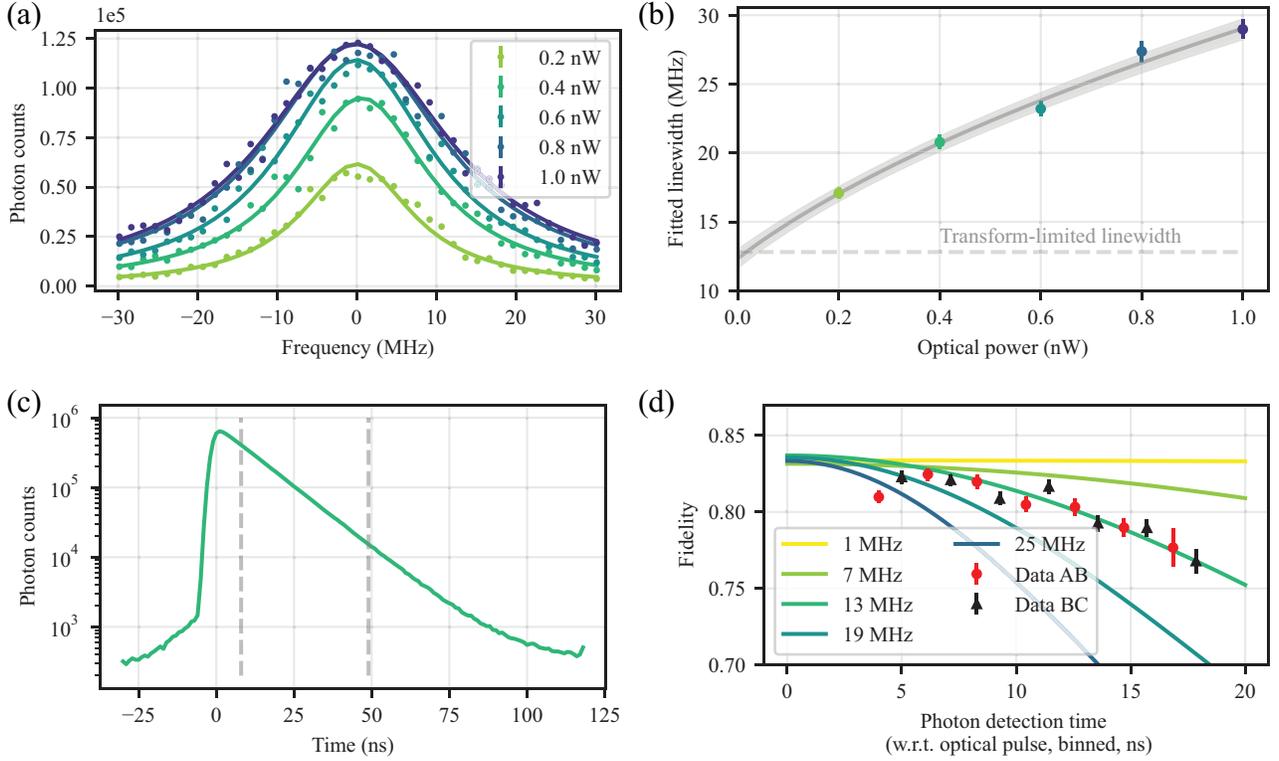}
	\caption{\textbf{Transform limited spectral linewidth.} \figlet{a} After passing a Charge-Resonance (CR) check, see Section \ref{sec:exp_setup}, we expose the emitter to a laser pulse with a variable frequency and count the emitted photons. We fit the result with a Lorentzian pulse shape (solid lines) and repeat the measurements for different powers of the applied laser pulse. \figlet{b} Extracted linewidths (colored circles) from panel (A) as function of optical power. We fit the curve given by \eref{lw_power_scaling} to extract the natural linewidth (solid gray line). The dashed gray line shows the expected transform-limited linewidth. \figlet{c} Histogram of phonon-side band photon (PSB) counts. We use the region between the dashed lines to extract the excited state lifetime and compute the expected linewidth in panel (B). \figlet{d} Fidelity with respect to the maximally entangled state as a function of the detection time of the heralding photon. We measure the entangled state fidelity for both the AB (red circles) and the BC (black triangles) links of Reference \cite{Pompili2021}. The detection time is binned in bins of 1 ns and the x value represents the start of the bin. The x axis is the detection time with respect to the highest intensity point of the optical pulse. Using a Monte-Carlo simulation, we model the entangled state fidelity for a frequency difference between the emitted photons given by a Gaussian distribution, for different values of the full-width half-maximum of the distribution.}
	\label{fig:photon_dist}
\end{figure*}

The next aspect we study is the photon distinguishability. For the case of indistinguishable photons, the which-path information is completely erased by the beam splitter. In turn, any distinguishability of the photons will therefore affect  their interference and hence alter the average heralded density matrix, see Section \ref{sec:sce_protocol}. A difference in arrival time or temporal shape of the photons will result in different probabilities to detect a photon originating from either node at a certain point in time (Section \ref{sec:bright_state_pop}). We note that different spatial modes of the photons will have a similar effect, but since, in our case, the photons are interfered on a fiber-coupled beam splitter this effect can be neglected. 

Distinguishability in polarization or frequency will act on the off-diagonal terms of the density matrix. A difference in polarization decreases the magnitude of the off-diagonal term, see \eref{off_diag_magnitude}. We work with a fiber-based beam splitter consisting of polarization maintaining (PM) fibers. We align the polarization to the slow or fast axis better than 20 dB, therefore we can assume the polarization mismatch to be small. As discussed in the previous section, a fixed frequency offset causes the entangled state phase to be dependent on the detection time of the photon. In Figure \ref{fig:ent_state_phase}e, we introduced a fixed frequency difference throughout the entire duration of the experiment. However, even if the average frequency offset is zero, if the frequency difference varies in each experimental run, an entangled state with a different phase will be heralded in each repetition. While averaging over many repetitions would still yield a constant phase, the fidelity with respect to the target state will be decreased, with lower fidelity for later detection times. 

In Section \ref{sec:exp_setup} we discussed the Charge-Resonance (CR) check as a way to ensure the NV centers are on resonance with the excitation laser and to eliminate any frequency difference between the emitters. Here we assess the performance of the CR check by measuring the spectral linewidth of the NV after passing the CR check. We turn on an additional laser with a variable frequency and count the emitted photons. We perform this procedure many times before moving to a different frequency setpoint of the additional laser. We scan over a range of $ \pm $ 30 MHz around the frequency of the excitation laser, see \figref{photon_dist}a. We fit the measured counts $N$ with a Lorentzian shape 
\begin{equation}\label{eq:lorentz}
    N = \frac{a}{\pi}\frac{2\gamma}{4f^2+\gamma^2} + A
\end{equation}
and extract $ \gamma $, the full-width at half-maximum (FWHM). In \eref{lorentz}, $a$ represents a scaling factor and $A$ an offset. Non-zero laser power induces broadening of the linewidth \cite{Citron1977}, so we repeat this measurement for different powers of the scanning laser. In \figref{photon_dist}b we plot the extracted linewidth for the different optical powers of the additional laser and fit the measured linewidths as a function of the optical power $ P $ with the expression
\begin{equation}\label{eq:lw_power_scaling}
\gamma = \sqrt{\gamma_0^2 + b\cdot P}
\end{equation}
to find the natural linewidth  $ \gamma_0 $. In \eref{lw_power_scaling}, $ b $ is the scaling factor relating the externally calibrated applied power to the optical Rabi-frequency. We find $ \gamma_0 =  (12.4 \pm 0.8)$ MHz and $ b = (690 \pm 40 )$ MHz$ ^2 $/nW. 

We compare the observed natural linewidth with the expected linewidth extracted from an excited-state lifetime measurement on the same NV center. We apply an optical $ \pi $-pulse to the NV center and record the detection times of the photons, see \figref{photon_dist}c. We fit the regime between the dashed lines, for which the influence of the pulse and dark counts is negligible, with an exponential decay and we find a lifetime $ \tau = (12.43 \pm 0.02) $ ns. The corresponding lifetime limited linewidth is $ \gamma_{0,l} = (12.81 \pm 0.02)$ MHz. We thus conclude that implementing a CR check can yield transform-limited linewidths within measurement accuracy and thus allows access to (near-)perfectly coherent photons. 

Having addressed the spectral properties of a single NV center, we now move to the photon distinguishability of two emitters and its effect on the entangled state fidelity. We measure the fidelity of generated entangled states between setups A and B as well as between setups B and C. In \figref{photon_dist}d, we show the measured fidelity (binned) versus the detection time of the photon. We observe a drop in fidelity for later detection times of the photon. Since the signal-to-noise ratio is approximately constant over the entire detection window, the observed drop in fidelity may be attributed to a varying frequency mismatch of the emitters. 

We use the Monte-Carlo simulation to predict how large the frequency mismatch would have to be to explain the measured behavior. We pick a random frequency difference from a Gaussian distribution, calculate the resulting entangled state phase and compute the fidelity to the target state (with target phase). By averaging over many repetitions and repeating for different widths of the Gaussian distribution, we obtain the entangled state fidelity as a function of detection time and FWHMs of the frequency distribution (solid lines in \figref{photon_dist}d). The measured entanglement data appears to be consistent with a fluctuating frequency mismatch with standard deviation of $ \approx 13 $ MHz between the emitters. Interestingly, we observe a quantitatively similar dependence for the link between Alice and Bob and for the link between Bob and Charlie. Note that this $\approx 13 $ MHz standard deviation of the frequency mismatch is inconsistent with the observed lifetime limited linewidth and the accuracy of the  wavemeter to which the lasers are locked ($<$2 MHz). Future work should focus on confirming the nature of this noise and its source.

\section{Double optical excitation} \label{sec:double_exc}

\begin{figure}
	\includegraphics{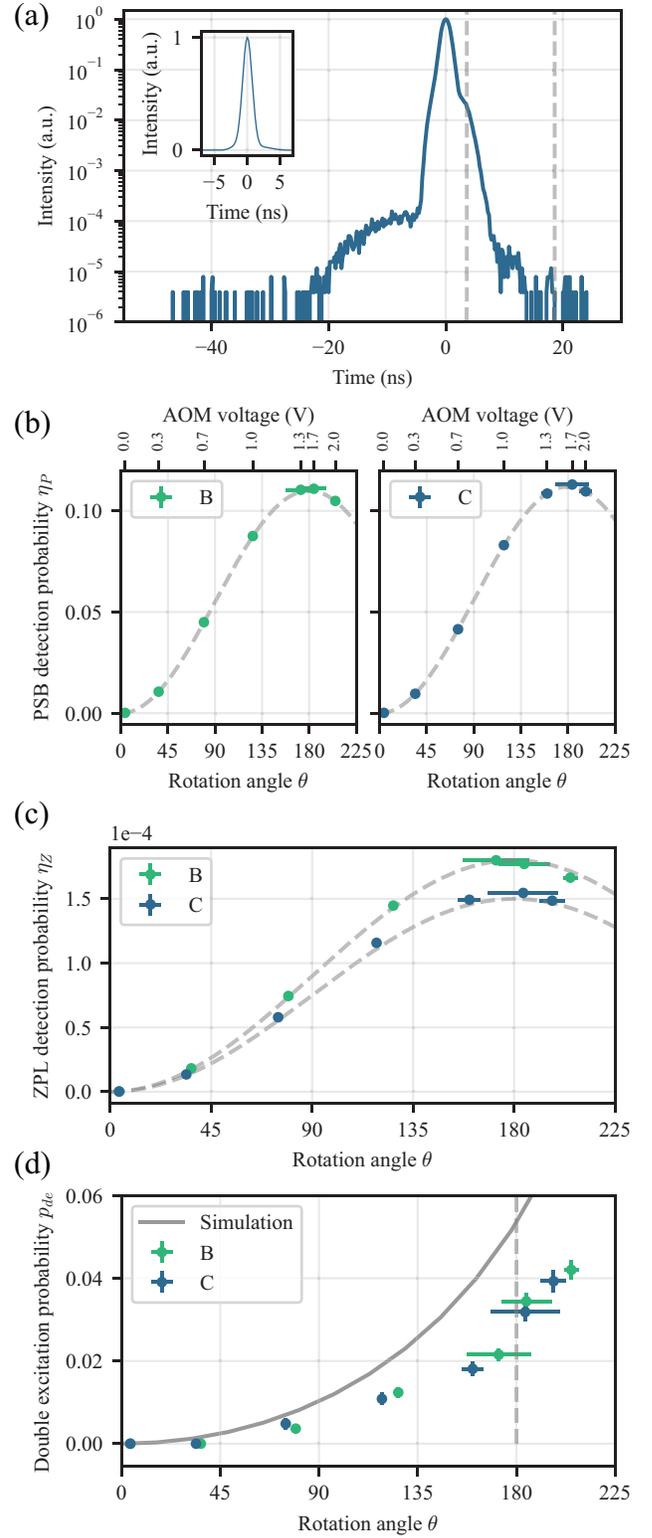}
	\caption{\textbf{Double excitation probability.} \figlet{a} Temporal shape of the excitation pulse. The exact shape is determined by the different components we use to generate the pulse. The region between the dashed lines is the detection window in which we accept heralding photons. The inset displays the temporal shape of the pulse on a linear scale. \figlet{b}\figlet{c} Photon detection probabilities for off-resonant phonon-side band (PSB) photons \figlet{b} and resonant zero-phonon line (ZPL) photons \figlet{c}. The dashed curves show expected behavior. In \figlet{b} we indicate the voltage applied to the AOM in the top axis. \figlet{d} Extracted double excitation probability $p_{de}$ using \eref{double_exc} and simulated curve resulting from our model (gray line). }
	\label{fig:double_exc}
\end{figure}
In Sections \ref{sec:sce_protocol} and \ref{sec:tailored_model} we have briefly discussed double optical excitation. During the finite duration of the excitation pulse, the emitter can get re-excited after emission of a first photon. In the high-photon loss regime, the probability that both photons arrive at the beam splitter is negligible. Loss of one of the two photons projects the qubits and result in a lowered fidelity of the heralded state. In our experiment we start the detection window a few nanoseconds after the highest intensity point of the pulse. We define double excitation as the probability that one emitter has emitted two photons given a photon detection event in the heralding window, $p_{de} = |\zeta(t_r,t)|^2$. Here we extract the double excitation probability for our specific optical excitation pulse and measure its dependence on the power of the optical pulse. 

The double excitation probability depends on the pulse duration with respect to the lifetime of the excited state, but also on the exact pulse shape and power of the pulse. As mentioned in Section \ref{sec:exp_setup}, we generate the optical pulse using an AWG, EOM and AOM. The combination of the response times and output of these three instruments determines the shape of the excitation pulse. In Figure \ref{fig:double_exc}a we plot the optical pulse intensity and indicate the heralding detection window.

We measure the double excitation probability for different powers of the optical pulse for nodes B and C. We scale the intensity of the pulse by changing the voltage sent to the AOM to maintain the same pulse shape for different powers (as opposed to changing the amplitude of the EOM pulse). We measure the emitted photons using two different detection paths, namely the detection path of the resonant ZPL photons (one of the detectors after the beam splitter in Figure \ref{fig:levels}a), and an additional detection path for the off-resonant PSB photons \cite{Hermans2021}. We extract the optical rotation angles $\theta$ and the detection efficiency $\eta_p$ ($\eta_z)$ from the photon detection probability in the PSB (ZPL) detector (Figure \ref{fig:double_exc}b), assuming that double excitation is small. The optical rotation angle $\theta$ serves as a measure of how much of the population of the $\ket{0}$ ground state is transferred to the excited state $\ket{e}$, where 180$^o$ marks the point where all population has been transferred. For each power, or rotation angle $\theta$, we measure the number of coincidence events where both detectors (ZPL and PSB) detected a photon, $N_{\text{coin}}$. The ZPL detection window is indicated in Figure \ref{fig:double_exc}a and we set the detection window of the PSB photons to start and end well before and after the pulse.

Since we start the heralding detection window of the ZPL after the optical pulse has (approximately) ended, re-excitation is not possible during this window, so we can assume that all measured coincidence events consist of a PSB photon during the optical pulse and a ZPL photon in the detection window. To compute the double excitation probability $p_{de}$ from the measured number of coincidence events $N_{\text{coin}}$, we can thus reformulate $p_{de}$ as  
\begin{equation}\label{eq:pde_exp}
	p_{de} = \frac{P_2}{P_1 +P_2},
\end{equation} 
where $P_1$ ($P_2$) is the probability of a single (two) photon emission, given a ZPL photon detected in the window. However, single or two photon emissions have different probabilities to be detected, and we can define the corresponding detection probabilities as
\begin{align}
	P_2' =& \eta_P \eta_Z P_2, \\ 
	P_1' = &\eta_Z P_1.
\end{align}
Here $\eta_Z$ and $\eta_P$ are the detection efficiencies of the ZPL and PSB detection paths respectively. Filling in the definitions of $P_1$ and $P_2$ in \eref{pde_exp} gives
\begin{align}\label{eq:double_exc}
\begin{split}
	p_{de} =& \frac{P_2'}{\eta_P P_1' + P_2'}, \\
	=& \frac{N_{\text{coin}}}{\eta_P N_{\text{all}} - (\eta_P-1)N_{\text{coin}}},
\end{split}
\end{align}
in which we have used $N_{\text{coin}} = n P_2'$ and $N_{\text{all}} = n (P_1' + P_2')$ for $n$ repetitions.

 Figure \ref{fig:double_exc}d shows the extracted double excitation probability $p_{de}$ for different rotation angles $ \theta $ for the two emitters (turquoise and blue circles), together with the simulated double excitation probability for our exact pulse shape using the model from Reference \cite{Hermans2021}\footnote{The numerical model and the exact pulse shape are included in the supporting software and data package\cite{software_and_data}}. There is a clear qualitative agreement between the data and the simulations. The quantitative difference between the data and simulations could potentially be explained by measurement errors in the pulse shape displayed in Figure \ref{fig:double_exc}a. We note that the simulations are very sensitive to the exact shape of the pulse, and any measurement artifact such as reflections in the optical path could broaden the measured pulse shape. Importantly, we find from both the measured data and simulations that the double optical excitation probability can be mitigated by choosing a smaller rotation angle $ \theta $, albeit at the cost of a lowered entanglement generation rate (which scales with $\cst$). In contrast, setting a $ \theta > 180^o $ results both in a lowered entanglement rate and fidelity, and therefore should be avoided.

\section{Non-excited bright state population}\label{sec:non_pi_pulse}

\begin{figure}
	\includegraphics{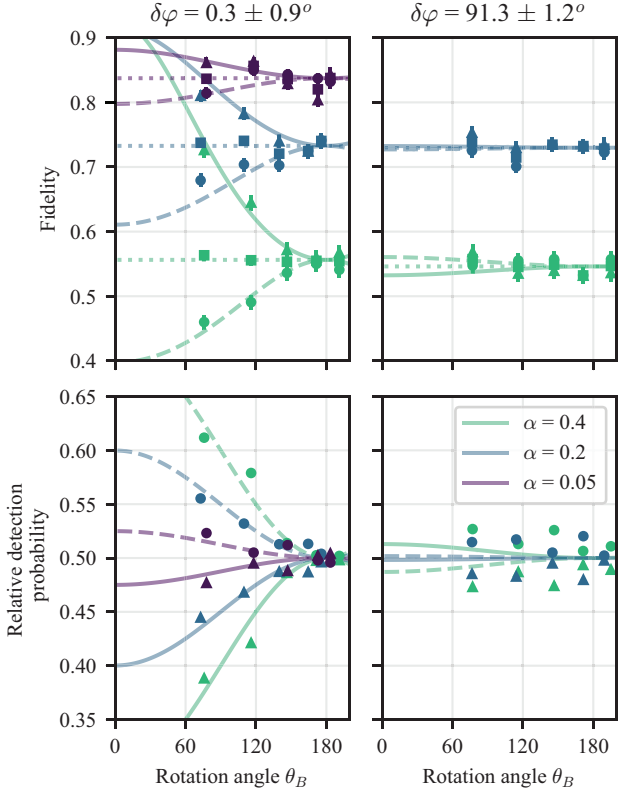}
	\caption{\textbf{Fidelity and relative detection probability in case of non-excited bright state population.} In the top panels, we measure the entangled state fidelities as a function of optical excitation rotation angle $ \theta $ for states heralded by a photon detection in detectors 1 (circles) and 2 (triangles), and the weighted average (squares) for different values of $ \alpha $ ( $ \alpha=0.05,0.2,0.4 $ for the purple, blue and green data points respectively). In the bottom panels we plot the relative probability to detect the heralding photon in each detector. We perform these measurements for two different setpoints $ \delta \varphi $ of the optical phase stabilization (left and right panels). The x-axis indicates the rotation angle of node B. The rotation angle of node C is approximately equal, but differs slightly due to small differences in the delivered optical power to the nodes. The dashed and solid lines are the results of the model (see main text). To include other error sources than the protocol error, we scale the results of the model to the measured average fidelities.}
	\label{fig:00_interference}
\end{figure}

Up to now we have considered perfect optical pulses, such that $c_{0,i} = 0$. In this section, we study the case of $c_{0,i} \neq 0$. To isolate the effect of $c_{0,i} \neq 0$, we make the following assumptions. We assume high photon loss ($\eta \ll 1$), no double excitation ($|c_{ee,i}|^2 = 0$), perfectly overlapping polarization of the photons ($\vec{\epsilon_A} \cdot \vec{\epsilon_B} = 1$), no frequency difference between the emitters ($\omega_A = \omega_B$), no noise photons ($p_d = 0$), we consider the two setups to have equal bright state populations and photon losses ($\alpha_A=\alpha_B\equiv\alpha, \eta_A= \eta_B\equiv\eta$) and we write the phase of the entangled state as the optical phase difference in front of the beam splitter $\phi = - \delta\varphi$. Furthermore, we define 
\begin{align}
c_0 =& \cos\frac{\theta}{2}\\
c_e =& \sin\frac{\theta}{2}, 
\end{align}
where $\theta$ can be considered as the optical excitation rotation angle between the ground and excitation state, as explained in the previous Section.

We rewrite Equations (\ref{eq:pclick}) and (\ref{eq:final_rho}) using these assumptions. The probability $p_{\text{click,C/D}}$ to detect a photon in either port C or D of the beam splitter is now given by 
\begin{equation}\label{eq:non_exc_pclick}
	\begin{split}
	p_{\text{click,C/D}} = &  \alpha \eta\sst \pm \alpha^2 \eta\cos\delta\varphi\sst\cst, \\
	= & \alpha \eta \sst (1 \pm \alpha\cos\delta\varphi\cst ),
	\end{split}
\end{equation}
where the +(-) sign correspond to a photon detection in port C (port D).
Using the corresponding density matrices $\rho_{C,D}$, we compute the fidelity with the maximally entangled state with a phase $\phi_T$ as
\begin{equation}\label{eq:non_exc_fid}
	\begin{split}
	F_{C/D} = & \bra{\Psi}\rho_{C,D}\ket{\Psi} \\
	= & \frac{1}{4p_{\text{click,C/D}}}(a_{11} + a_{22} + a_{12}e^{-i\phi_T} + a_{12}^\ast e^{i\phi_T}) \\
	= & \frac{(1-\alpha)}{2(1 \pm \alpha \cos \delta\varphi \cst)}(1+\cos(\delta\varphi - \phi_T))
	\end{split}
\end{equation}

From Equations (\ref{eq:non_exc_pclick}) and (\ref{eq:non_exc_fid}) it is apparent that the fidelity of the heralded entangled state depends on which detector detects the single photon, the optical phase difference $ \delta \varphi $ and the optical excitation rotation angle $ \theta $. Surprisingly, for $ \delta \varphi = 0 $ and $ \theta \rightarrow 0 $ the fidelity of the entangled state heralded by one of the detectors approaches 1, albeit with a small probability to occur. On the contrary, for $ \delta \varphi = 90^o $ no difference in entangled state fidelities and their probabilities to be heralded is expected. This result can be interpreted as the interference of the different photonic states associated with the $\ket{00}_{AB}$ qubit states. Dependent on $\delta \varphi$, constructive or destructive interference of the $\hat{a}_A^\dagger\ket{0}_\gamma\ket{00}_{AB} $ state with $\hat{a}_B^\dagger\ket{0}_\gamma\ket{00}_{AB} $ causes different heralding probabilities and consequently different average heralded state fidelities. The fact that the heralded fidelity can approach 1 for particular detector and settings is explained by the reduction of the protocol error (the error resulting from both qubits being in the bright state): a low excitation probability makes the probability that two photons were emitted small and destructive interference ensures that if one photon was emitted it is directed towards the other detector.

\begin{figure}
    \centering
    \includegraphics{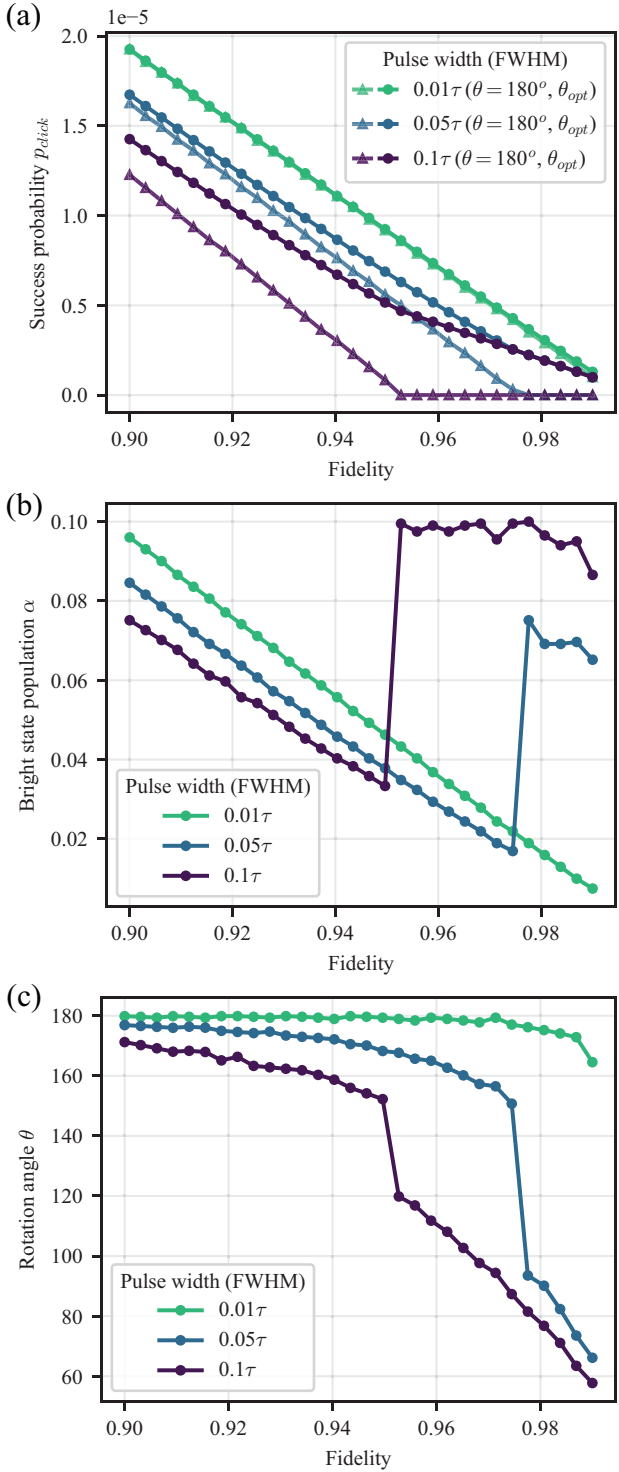}
    \caption{\textbf{Entangled state optimization.} \figlet{a} For a fixed fidelity with respect to the maximally entangled state, we extract the maximal success probability $p_{\text{click}}$ for different temporal widths of the excitation pulse, by varying $\alpha$ and fixing the optical rotation angle $\theta = 180^o$ (circles) or by varying $\alpha$ and $\theta$ (triangles).Note that for the smallest pulse width (green line) both optimization nearly overlap. \figlet{b} \figlet{c} For the optimization where both $\alpha$ and $\theta$ are varied (circles in panel \figlet{a}), the extracted optimal settings for the bright state population $\alpha$ \figlet{b} and the optical excitation rotation angle $\theta$ \figlet{c} are shown.} 
    \label{fig:fid_opt}
\end{figure}

We compare these theoretical results with our experimental data in Figure \ref{fig:00_interference}. We stabilize the interferometer to different setpoints to obtain $ \delta \varphi  = 0.3 \pm 0.9^o$ and $ \delta \varphi  = 91.3 \pm 1.2^o$ (left and right panels). For different values of $ \alpha $ ($ \alpha = [0.05, 0.2, 0.4] $, purple, blue and green data points respectively) we generate entanglement while varying the optical rotation angle $\theta$. We record the fidelity (top panels) for heralding signals detected by the two different detectors (triangles and circles) and the mean fidelity (square data points), and the relative probability to detect a photon on each detector (bottom panels). In the same figure we plot the theoretical model scaled to the mean measured fidelity to incorporate additional errors (solid, dashed and dotted lines). Our measured data is in excellent agreement with the theoretical model, we observe the effect of the lowered excitation power on the fidelity heralded by the different detectors for the case $ \delta \varphi \approx 0^o $ and the absence of this effect for $ \delta \varphi \approx 90^o $. Note that the (mean) fidelity for small rotation angles $\theta$ is additionally improved by the reduction of the double excitation errors due to the lowered optical power, see Figure \ref{fig:double_exc}d. However the data displayed here has a too high measurement uncertainty to resolve this effect (we expect a maximum improvement of the fidelity $ F $ by $ \approx 0.01$ for the measured rotation angles).

\section{State optimization}\label{sec:opt}

Let us now turn again to the optimization of the entanglement generation. The effects discussed in the previous two sections, double optical excitation and non-excited bright state population, are both related to the optical power of the excitation pulse. Together with the bright state population (as discussed in Section \ref{sec:bright_state_pop}), we have two degrees of freedom in optimizing the entangled state and we would like to find the optimal settings to produce a state with fixed fidelity with the highest success probability. In this Section, we assume symmetric settings of the experimental setups ($\alpha_A = \alpha_B, \eta_A = \eta_B, \theta_A = \theta_B$). Furthermore, we assume no noise counts and perfectly indistinguishable photons.

In Figure \ref{fig:fid_opt}a, we plot the maximal success probability versus fidelity of the (numerically simulated) heralded state for different temporal widths of the excitation pulse (estimated as a Gaussian shaped pulse), in units of the excited state lifetime $\tau$. The triangular data points represent the case where $\alpha$ is varied, but the optical power of the excitation pulse is set such that it transfers all population from the ground to the excited state (rotation angle $\theta = 180^o$). The circular data points show the case where both $\alpha$ and $\theta$ are varied. Figure \ref{fig:fid_opt}b and \ref{fig:fid_opt}c show the settings for the bright state population $\alpha$ and optical excitation angle $\theta$ that optimize the success probability for fixed fidelity.

We can interpret the strategy for the optimized settings as follows. When the intrinsic protocol error ($\alpha$) is the dominating source of infidelity, it is most efficient to choose a smaller $\alpha$ to increase the entangled state fidelity. Furthermore, the numerical optimization reproduces our conclusion from Section \ref{sec:double_exc}; reducing slightly the optical power of the excitation pulse mitigates the effect of double excitation at only a small cost in success probability. On the contrary, when the double excitation contribution becomes the dominating error source, lowering the optical power allows the target fidelity to be met while with fixed $\theta$ these fidelities would be impossible to reach, which can be seen from the sudden change of behavior in Figures \ref{fig:fid_opt}b and c.

\section{Conclusions and discussion}\label{sec:conclusion}

In conclusion, we have performed a detailed theoretical and experimental investigation of the single-photon entanglement protocol.  We have developed a general model for states heralded using the single-photon protocol, tailored the model to our experimental setting, NV centers in bulk diamond, and experimentally verified the effect of several experimental parameters.

 We have studied the effect of the bright state population $ \alpha $ on the generated entangled state and the success probability to herald a state. We have demonstrated the entangled state phase dependence on MW pulse phases $ \vartheta $, the optical phase stabilization setpoint $ \delta \varphi $ and the detection time of the heralding photon detection in combination with an emission frequency difference between the qubits. We have shown the observation of transform-limited spectral linewidths, using a Charge-Resonance (CR) check to remove any spectral shift. Nonetheless, our data on remote entanglement is consistent with a residual fluctuating frequency shift between the emitters, this will be a subject for future work. We have observed a decrease of the double excitation probability for lowered optical power of the excitation laser pulses. Additionally, we have shown that reducing the optical laser power can also lead to different heralded state fidelities dependent on which detector heralded the state. Lastly, we have shown that the optical laser power, together with the initial bright state population, can be used to optimize the state fidelity and success probability.

Finally, while the experiments carried out in this work involved the nitrogen-vacancy center as a qubit platform, the  conclusions presented here are readily applicable to other qubit platforms, such as other solid state defects and quantum dots. The insights gained in this work will be crucial in improving the entangled state fidelities using the single-photon entanglement protocol.

\bibliographystyle{ieeetr} 
\bibliography{bibliography}

\begin{thebibliography}{10}

\bibitem{Kimble2008}
H.~J. Kimble, ``{The quantum internet},'' {\em Nature}, vol.~453, no.~7198,
  pp.~1023--1030, 2008.

\bibitem{Wehner2018}
S.~Wehner, D.~Elkouss, and R.~Hanson, ``{Quantum internet: A vision for the
  road ahead},'' {\em Science}, vol.~362, no.~6412, 2018.

\bibitem{Ekert2014}
A.~Ekert and R.~Renner, ``{The ultimate physical limits of privacy},'' {\em
  Nature}, vol.~507, no.~7493, pp.~443--447, 2014.

\bibitem{Broadbent2009}
A.~Broadbent, J.~Fitzsimons, and E.~Kashefi, ``{Universal blind quantum
  computation},'' {\em Proceedings - Annual IEEE Symposium on Foundations of
  Computer Science, FOCS}, pp.~517--526, 2009.

\bibitem{Ben-Or2006}
M.~Ben-Or, C.~Cr{\'{e}}peau, D.~Gottesman, A.~Hassidim, and A.~Smith, ``{Secure
  multiparty quantum computation with (only) a strict honest majority},'' {\em
  Proceedings - Annual IEEE Symposium on Foundations of Computer Science,
  FOCS}, pp.~249--258, 2006.

\bibitem{Christandl2005}
M.~Christandl and S.~Wehner, ``{Quantum anonymous transmissions},'' {\em
  Lecture Notes in Computer Science (including subseries Lecture Notes in
  Artificial Intelligence and Lecture Notes in Bioinformatics)}, vol.~3788
  LNCS, pp.~217--235, 2005.

\bibitem{DeBone2020}
S.~W. de~Bone, R.~Ouyang, K.~Goodenough, and D.~Elkouss, ``{Protocols for
  Creating and Distilling Multipartite GHZ States With Bell Pairs},'' in {\em
  IEEE Quantum Engineering}, 2020.

\bibitem{Cabrillo1999}
C.~Cabrillo, J.~I. Cirac, P.~Garcia-Fernandez, and P.~Zoller, ``{Creation of
  entangled states of distant atoms by interference},'' {\em Physical Review
  A}, vol.~59, no.~2, pp.~1025--1033, 1999.

\bibitem{Bose1999}
S.~Bose, P.~L. Knight, M.~B. Plenio, and V.~Vedral, ``{Proposal for
  teleportation of an atomic state via cavity decay},'' {\em Physical Review
  Letters}, vol.~83, no.~24, pp.~5158--5161, 1999.

\bibitem{Barrett2005}
S.~D. Barrett and P.~Kok, ``{Efficient high-fidelity quantum computation using
  matter qubits and linear optics},'' {\em Physical Review A - Atomic,
  Molecular, and Optical Physics}, vol.~71, no.~6, pp.~2--5, 2005.

\bibitem{Delteil2016}
A.~Delteil, Z.~Sun, W.~B. Gao, E.~Togan, S.~Faelt, and A.~Imamoglu,
  ``{Generation of heralded entanglement between distant hole spins},'' {\em
  Nature Physics}, vol.~12, no.~3, pp.~218--223, 2016.

\bibitem{Stockill2017}
R.~Stockill, M.~J. Stanley, L.~Huthmacher, E.~Clarke, M.~Hugues, A.~J. Miller,
  C.~Matthiesen, C.~{Le Gall}, and M.~Atat{\"{u}}re, ``{Phase-Tuned Entangled
  State Generation between Distant Spin Qubits},'' {\em Physical Review
  Letters}, vol.~119, no.~1, pp.~1--6, 2017.

\bibitem{Humphreys2018}
P.~C. Humphreys, N.~Kalb, J.~P. Morits, R.~N. Schouten, R.~F. Vermeulen, D.~J.
  Twitchen, M.~Markham, and R.~Hanson, ``{Deterministic delivery of remote
  entanglement on a quantum network},'' {\em Nature}, vol.~558, no.~7709,
  pp.~268--273, 2018.

\bibitem{Lago-Rivera2021}
D.~Lago-Rivera, S.~Grandi, J.~V. Rakonjac, A.~Seri, and H.~de~Riedmatten,
  ``{Telecom-heralded entanglement between remote multimode solid-state quantum
  memories},'' {\em Nature}, vol.~594, no.~June, 2021.

\bibitem{rose_observation_2018}
B.~C. Rose, D.~Huang, Z.-H. Zhang, P.~Stevenson, A.~M. Tyryshkin,
  S.~Sangtawesin, S.~Srinivasan, L.~Loudin, M.~L. Markham, A.~M. Edmonds, D.~J.
  Twitchen, S.~A. Lyon, and N.~P. de~Leon, ``Observation of an environmentally
  insensitive solid-state spin defect in diamond,'' {\em Science}, vol.~361,
  pp.~60--63, July 2018.
\newblock Publisher: American Association for the Advancement of Science.

\bibitem{bhaskar_quantum_2017}
M.~Bhaskar, D.~Sukachev, A.~Sipahigil, R.~Evans, M.~Burek, C.~Nguyen,
  L.~Rogers, P.~Siyushev, M.~Metsch, H.~Park, F.~Jelezko, M.~Lončar, and
  M.~Lukin, ``Quantum {Nonlinear} {Optics} with a {Germanium}-{Vacancy} {Color}
  {Center} in a {Nanoscale} {Diamond} {Waveguide},'' {\em Physical Review
  Letters}, vol.~118, p.~223603, May 2017.
\newblock Publisher: American Physical Society.

\bibitem{nguyen_quantum_2019}
C.~Nguyen, D.~Sukachev, M.~Bhaskar, B.~Machielse, D.~Levonian, E.~Knall,
  P.~Stroganov, R.~Riedinger, H.~Park, M.~Lončar, and M.~Lukin, ``Quantum
  {Network} {Nodes} {Based} on {Diamond} {Qubits} with an {Efficient}
  {Nanophotonic} {Interface},'' {\em Physical Review Letters}, vol.~123,
  p.~183602, Oct. 2019.
\newblock Publisher: American Physical Society.

\bibitem{trusheim_transform-limited_2020}
M.~E. Trusheim, B.~Pingault, N.~H. Wan, M.~Gündoğan, L.~De~Santis,
  R.~Debroux, D.~Gangloff, C.~Purser, K.~C. Chen, M.~Walsh, J.~J. Rose, J.~N.
  Becker, B.~Lienhard, E.~Bersin, I.~Paradeisanos, G.~Wang, D.~Lyzwa, A.~R.-P.
  Montblanch, G.~Malladi, H.~Bakhru, A.~C. Ferrari, I.~A. Walmsley,
  M.~Atatüre, and D.~Englund, ``Transform-{Limited} {Photons} {From} a
  {Coherent} {Tin}-{Vacancy} {Spin} in {Diamond},'' {\em Physical Review
  Letters}, vol.~124, p.~023602, Jan. 2020.
\newblock Publisher: American Physical Society.

\bibitem{widmann_coherent_2015}
M.~Widmann, S.-Y. Lee, T.~Rendler, N.~T. Son, H.~Fedder, S.~Paik, L.-P. Yang,
  N.~Zhao, S.~Yang, I.~Booker, A.~Denisenko, M.~Jamali, S.~A. Momenzadeh,
  I.~Gerhardt, T.~Ohshima, A.~Gali, E.~Janzén, and J.~Wrachtrup, ``Coherent
  control of single spins in silicon carbide at room temperature,'' {\em Nature
  Materials}, vol.~14, pp.~164--168, Feb. 2015.
\newblock Number: 2 Publisher: Nature Publishing Group.

\bibitem{son_developing_2020}
N.~T. Son, C.~P. Anderson, A.~Bourassa, K.~C. Miao, C.~Babin, M.~Widmann,
  M.~Niethammer, J.~Ul~Hassan, N.~Morioka, I.~G. Ivanov, F.~Kaiser,
  J.~Wrachtrup, and D.~D. Awschalom, ``Developing silicon carbide for quantum
  spintronics,'' {\em Applied Physics Letters}, vol.~116, p.~190501, May 2020.
\newblock Publisher: American Institute of Physics.

\bibitem{dibos_atomic_2018}
A.~Dibos, M.~Raha, C.~Phenicie, and J.~Thompson, ``Atomic {Source} of {Single}
  {Photons} in the {Telecom} {Band},'' {\em Physical Review Letters}, vol.~120,
  p.~243601, June 2018.
\newblock Publisher: American Physical Society.

\bibitem{kindem_control_2020}
J.~M. Kindem, A.~Ruskuc, J.~G. Bartholomew, J.~Rochman, Y.~Q. Huan, and
  A.~Faraon, ``Control and single-shot readout of an ion embedded in a
  nanophotonic cavity,'' {\em Nature}, vol.~580, pp.~201--204, Apr. 2020.

\bibitem{coste_high-rate_2022}
N.~Coste, D.~Fioretto, N.~Belabas, S.~C. Wein, P.~Hilaire, R.~Frantzeskakis,
  M.~Gundin, B.~Goes, N.~Somaschi, M.~Morassi, A.~Lemaître, .~I. Sagnes,
  A.~Harouri, S.~E. Economou, A.~Auffeves, O.~Krebs, L.~Lanco, and
  P.~Senellart, ``High-rate entanglement between a semiconductor spin and
  indistinguishable photons,'' July 2022.
\newblock arXiv:2207.09881 [quant-ph].

\bibitem{Das2019}
S.~Das, L.~Zhai, M.~{\v{C}}epulskovskis, A.~Javadi, S.~Mahmoodian, P.~Lodahl,
  and A.~S. S{\o}rensen, ``{A wave-function ansatz method for calculating field
  correlations and its application to the study of spectral filtering and
  quantum dynamics of multi-emitter systems},'' {\em arXiv:1912.08303}, 2019.

\bibitem{tiurev_fidelity_2021}
K.~Tiurev, P.~L. Mirambell, M.~B. Lauritzen, M.~H. Appel, A.~Tiranov,
  P.~Lodahl, and A.~S. Sørensen, ``Fidelity of time-bin-entangled multiphoton
  states from a quantum emitter,'' {\em Physical Review A}, vol.~104, no.~5,
  p.~052604, 2021.
\newblock Publisher: American Physical Society.

\bibitem{Robledo2011a}
L.~Robledo, H.~Bernien, T.~v.~d. Sar, and R.~Hanson, ``Spin dynamics in the
  optical cycle of single nitrogen-vacancy centres in diamond,'' {\em New
  Journal of Physics}, vol.~13, no.~2, p.~025013, 2011.

\bibitem{Robledo2011}
L.~Robledo, L.~Childress, H.~Bernien, B.~Hensen, P.~F.~A. Alkemade, and
  R.~Hanson, ``High-fidelity projective read-out of a solid-state spin quantum
  register,'' {\em Nature}, vol.~477, no.~7366, pp.~574--578, 2011.

\bibitem{Riedel2017}
D.~Riedel, I.~S{\"{o}}llner, B.~J. Shields, S.~Starosielec, P.~Appel, E.~Neu,
  P.~Maletinsky, and R.~J. Warburton, ``{Deterministic enhancement of coherent
  photon generation from a nitrogen-vacancy center in ultrapure diamond},''
  {\em Physical Review X}, vol.~7, no.~3, pp.~1--8, 2017.

\bibitem{Pompili2021}
M.~Pompili, S.~L. Hermans, S.~Baier, H.~K. Beukers, P.~C. Humphreys, R.~N.
  Schouten, R.~F. Vermeulen, M.~J. Tiggelman, L.~{dos Santos Martins},
  B.~Dirkse, S.~Wehner, and R.~Hanson, ``{Realization of a multinode quantum
  network of remote solid-state qubits},'' {\em Science}, vol.~372, no.~6539,
  pp.~259--264, 2021.

\bibitem{Hermans2021}
S.~L.~N. Hermans, M.~Pompili, H.~K.~C. Beukers, S.~Baier, J.~Borregaard, and
  R.~Hanson, ``Qubit teleportation between non-neighbouring nodes in a quantum
  network,'' {\em Nature}, vol.~605, no.~7911, pp.~663--668, 2022.

\bibitem{Tamarat2006}
P.~Tamarat, T.~Gaebel, J.~R. Rabeau, M.~Khan, A.~D. Greentree, H.~Wilson, L.~C.
  Hollenberg, S.~Prawer, P.~Hemmer, F.~Jelezko, and J.~Wrachtrup, ``{Stark
  shift control of single optical centers in diamond},'' {\em Physical Review
  Letters}, vol.~97, no.~8, pp.~1--4, 2006.

\bibitem{robledo_control_2010}
L.~Robledo, H.~Bernien, I.~van Weperen, and R.~Hanson, ``Control and
  {Coherence} of the {Optical} {Transition} of {Single} {Nitrogen} {Vacancy}
  {Centers} in {Diamond},'' {\em Physical Review Letters}, vol.~105, no.~17,
  p.~177403, 2010.

\bibitem{Doherty2011}
M.~W. Doherty, N.~B. Manson, P.~Delaney, and L.~C.~L. Hollenberg, ``The
  negatively charged nitrogen-vacancy centre in diamond: the electronic
  solution,'' {\em New Journal of Physics}, vol.~13, no.~2, p.~025019, 2011.

\bibitem{orphal-kobin_optically_2022}
L.~Orphal-Kobin, K.~Unterguggenberger, T.~Pregnolato, N.~Kemf, M.~Matalla,
  R.-S. Unger, I.~Ostermay, G.~Pieplow, and T.~Schröder, ``Optically coherent
  nitrogen-vacancy defect centers in diamond nanostructures,'' {\em arXiv:
  2203.05605}, 2022.

\bibitem{mccullian_quantifying_2022}
B.~A. McCullian, H.~F.~H. Cheung, H.~Y. Chen, and G.~D. Fuchs, ``Quantifying
  {NV}-center {Spectral} {Diffusion} by {Symmetry},'' {\em arXiv: 2206.11362},
  2022.

\bibitem{software_and_data}
S.~L.~N. Hermans, M.~Pompili, L.~Dos Santos~Martins, A.~R.-P. Montblanch,
  H.~K.~C. Beukers, S.~, Baier, J.~Borregaard, and R.~Hanson, ``The datasets
  supporting this manuscript and the software to analyze them will be made
  avaible soon..''

\bibitem{zhao_entangling_2014}
T.-M. Zhao, H.~Zhang, J.~Yang, Z.-R. Sang, X.~Jiang, X.-H. Bao, and J.-W. Pan,
  ``Entangling {Different}-{Color} {Photons} via {Time}-{Resolved}
  {Measurement} and {Active} {Feed} {Forward},'' {\em Physical Review Letters},
  vol.~112, p.~103602, Mar. 2014.
\newblock Publisher: American Physical Society.

\bibitem{vittorini_entanglement_2014}
G.~Vittorini, D.~Hucul, I.~V. Inlek, C.~Crocker, and C.~Monroe, ``Entanglement
  of distinguishable quantum memories,'' {\em Physical Review A}, vol.~90,
  p.~040302, Oct. 2014.
\newblock Publisher: American Physical Society.

\bibitem{Citron1977}
M.~L. Citron, H.~R. Gray, C.~W. Gabel, and C.~R. Stroud, ``{Experimental study
  of power broadening in a two-level atom},'' {\em Physical Review A}, vol.~16,
  no.~4, pp.~1507--1512, 1977.

\end{thebibliography}

\section*{Acknowledgments}
We acknowledge financial support from the EU Flagship on Quantum Technologies through the project Quantum Internet Alliance (EU Horizon 2020, grant agreement no. 820445); from the European Research Council (ERC) through an ERC Consolidator Grant (grant agreement no. 772627 to R.H.); from the Netherlands Organisation for Scientific Research (NWO) through a VICI grant (project no. 680-47-624) and the Zwaartekracht program Quantum Software Consortium (project no. 024.003.037/3368). S.B. acknowledges support from an Erwin-Schrödinger fellowship (QuantNet, no. J 4229-N27) of the Austrian National Science Foundation (FWF).

\clearpage
\onecolumngrid
\begin{appendices}
	\renewcommand{\thefigure}{A\arabic{figure}}
	\renewcommand{\theequation}{A.\arabic{equation}}
	\renewcommand{\thetable}{A\arabic{table}}
	\setcounter{figure}{1}
	\setcounter{equation}{1}
	\setcounter{table}{1}
\section{Average heralded density matrix}\label{app:full_state}

In this work we derive a general theoretical model for two-qubit states heralded by the single-photon entanglement protocol. The different steps and the corresponding unitaries are given in the main text. In this Appendix we write the expressions for the resulting density matrices for the different detection patterns. 

The average heralded density matrix for a photon detection in port C of the beam splitter is given by 
\begin{equation}\label{eq:final_rho_app}
\begin{split}
\rho_C = &\frac{1}{p_{\text{click,C}}} (\rho_{1} + \rho_{2} +\rho_{\text{incoherent}} + \rho_{\text{noise}})\\
\end{split}
\end{equation}
with the success probability $ p_{\text{click,C}} $
\begin{equation}\label{eq:pclick_app}
\begin{split}
p_{\text{click,C}} = & \Tr(\rho_{1}) +\Tr(\rho_{2}) +\Tr(\rho_{\text{incoherent}})+\Tr(\rho_{\text{noise}}).\\
\end{split}
\end{equation}

\noindent\textbf{Single photon}
In the case of a single detected photon and no loss, the density matrix is given by
\begin{equation}
\begin{split}
\rho_{1} =& \ket{\Psi_{4,1}} \bra{\Psi_{4,1}} \\
=& \frac{1}{2}
\begin{pmatrix}
a_{00} & a_{01} & a_{02} & 0 \\
a_{01}^\ast & a_{11}  & a_{12}   & 0 \\
a_{02}^\ast & a_{12}^\ast & a_{22} & 0 \\
0 & 0 & 0 & 0 \\
\end{pmatrix}
\end{split}
\end{equation}
with elements
\begin{align}
\begin{split}
a_{00} = & \alpha_A \alpha_B (c_{0,A}^2 \eta_B |\zeb|^2 + c_{0,B}^2 \eta_A |\zea|^2 + c_{0,A} c_{0,B} \sea \seb (\zeb\zea^\ast + \zea\zeb^\ast)) 
\end{split}
\\
\begin{split}
a_{01} = & \alpha_A \sab \smab e^{i\vartheta_B} (c_{0,A} \seb \zea^\ast \zeb + c_{0,B} \sea |\zea|^2)
\end{split}
\\
\begin{split}
a_{02} = &  \saa \smaa \alpha_B e^{i\vartheta_A} (c_{0,A} \seb |\zeb|^2  + c_{0,B} \sea \zea \zeb^\ast)
\end{split}
\\
\begin{split}
a_{11} = &  \alpha_A (1-\alpha_B) \eta_A |\zea|^2
\end{split}
\\
\begin{split}
a_{12} = &  \saa \smaa \sab \smab \sea \seb e^{-i(\vartheta_B - \vartheta_A)} \zea\zeb^\ast
\end{split}
\\
\begin{split}
a_{22} = &   (1-\alpha_A) \alpha_B \eta_B |\zeb|^2
\end{split}
\end{align}
\\
Here we use a short-hand notation for the photonic modes as defined in \eref{det_mode}, where$\zeta_i$ represents the detected mode at time $t'$.

\noindent\textbf{Two photons} 
When two photons arrive at the beam splitter without any lost photon, we accept an heralding event when both photons are being detected in port C of the beam splitter because we assume non-number resolving detectors. The first photon is detected at time $t'$ and the second photon at time $t''$. For these states we obtain the following density matrix
\begin{equation}
\begin{split}
\rho_{2} =& \ket{\Psi_{4,2}} \bra{\Psi_{4,2}} \\
=& \frac{1}{4}
\begin{pmatrix}
a_{00} & a_{01} & a_{02} & 0 \\
a_{01}^\ast & a_{11}  & a_{12}   & 0 \\
a_{02}^\ast & a_{12}^\ast & a_{22} & 0 \\
0 & 0 & 0 & 0 \\
\end{pmatrix}
\end{split}
\end{equation}
with elements 
\begin{align}
\begin{split}
a_{00} = & \alpha_A\alpha_B (c_{0,B}^2 \eta_A^2 |\zeaa|^2 + c_{0,A}^2 \eta_B^2 |\zebb|^2 + c_{0,A} c_{0,B} \eta_A \eta_B (\zeaa\zebb^\ast +\zebb\zeaa^\ast) \\
& + \eta_A\eta_B(|\zea(t')\zeb(t'') + \zea(t'')\zeb(t')|^2) \\
& + c_{0,A}\sea \eta_B\seb (\zebb\zea^\ast(t')\zeb^\ast(t'') + \zebb\zea^\ast(t'')\zeb^\ast(t') + \zea(t')\zeb(t'')\zebb^\ast +\zea(t'')\zeb(t')\zebb^\ast ) \\
& + c_{0,B} \eta_A\sea \seb (\zeaa\zea^\ast(t')\zeb^\ast(t'') + \zeaa\zea^\ast(t'')\zeb^\ast(t') + \zea(t')\zeb(t'')\zeaa^\ast +\zea(t'')\zeb(t')\zeaa^\ast )) 
\end{split}
\\
\begin{split}
a_{01} = & \alpha_A \sab \smab e^{i\vartheta_B} \\
& (c_{0,A}\eta_A\eta_B\zeaa^\ast\zebb +\eta_A\sea \seb \zeaa^\ast(\zea(t')\zeb(t'')+ \zea(t'')\zeb(t')) + \eta_A^2 |\zeaa|^2)
\end{split}
\\
\begin{split}
a_{02} = &  \saa \smaa \alpha_B e^{i\vartheta_A} \\
& (c_{0,B}\eta_A\eta_B\zebb^\ast\zeaa +\sea \eta_B\seb \zebb^\ast(\zea(t')\zeb(t'')+ \zea(t'')\zeb(t')) + \eta_B^2 |\zebb|^2)  
\end{split}
\\
\begin{split}
a_{11} = &  \alpha_A(1-\alpha_B) \eta_A^2 |\zeaa|^2  
\end{split}
\\
\begin{split}
a_{12} = &  \saa \smaa \sab \smab \eta_A \eta_B e^{-i(\vartheta_B-\vartheta_A)}\zeaa\zebb^\ast
\end{split}
\\
\begin{split}
a_{22} = &  (1-\alpha_A)\alpha_B \eta_B^2 |\zebb|^2
\end{split}
\end{align}
Again, here we have used a short hand notation for the photonic modes. The joint modes $\zeta_{ii}$ denote the detected modes at time $t'$ and $t''$.

\noindent\textbf{Lost photons} We consider all the detection patterns for which at least one photon is being detected in port C of the beam splitter and at least one photon is lost. We sum over all the individual detection patterns and arrive at
\begin{equation}
\begin{split}
\rho_{\text{incoherent}} =  &  \sum_{i=1,2} \ket{\Psi_{4,i,r}}\bra{\Psi_{4,i,r}} \\
=& \frac{1}{2}
\begin{pmatrix}
a_{00} & 0 & 0 & 0 \\
0 & a_{11}  & 0  & 0 \\
0 & 0& a_{22} & 0 \\
0 & 0 & 0 & 0 \\
\end{pmatrix}
\end{split}
\end{equation}
with elements
\begin{align}
\begin{split}
a_{00} = & \alpha_A \alpha_B ( \\
& \eta_A |\zea(t')|^2 ((1-\eta_B)|\zeb(t_r)|^2) + c_{ee,B}^2 - \eta_B^2 |\zebb(t'',t''')|^2)\\
& + \eta_B |\zeb(t')|^2 ((1-\eta_A)|\zea(t_r)|^2) + c_{ee,A}^2 - \eta_A^2 |\zeaa(t'',t''')|^2)\\
& + \eta_A(1-\eta_A)|\zeaa(t',t_r)|^2(c_{0,B}^2 + c_{e,B}^2 + c_{ee,B}^2) \\
& + \eta_A(1-\eta_A)|\zeaa(t_r,t')|^2(c_{0,B}^2 + c_{e,B}^2 + c_{ee,B}^2) \\
& + \eta_B(1-\eta_B)|\zebb(t',t_r)|^2(c_{0,A}^2 + c_{e,A}^2 + c_{ee,A}^2) \\
& + \eta_B(1-\eta_B)|\zebb(t_r,t')|^2(c_{0,A}^2 + c_{e,A}^2 + c_{ee,A}^2) \\
& + \eta_A^2 |\zeaa(t', t'')|^2 ( |\zeb(t_r)|^2 + c_{ee,B}^2 - \eta_B^2 |\zebb(t''',t'''')|^2)\\
& + \eta_B^2 |\zebb(t', t'')|^2 (|\zea(t_r)|^2 + c_{ee,A}^2 - \eta_A^2 |\zeaa(t''',t'''')|^2))\\
\end{split}
\\
\begin{split}
a_{11} = & \alpha_A (1-\alpha_B)\eta_A (1-\eta_A) (|\zeaa(t', t_r)|^2+|\zeaa(t_r, t')|^2)\\
\end{split}
\\
\begin{split}
a_{22} = &  (1-\alpha_A)\alpha_B \eta_B (1-\eta_B) (|\zebb(t', t_r)|^2 + |\zebb(t_r, t')|^2) \\
\end{split}
\end{align}
Here we have used the relations $\eta_i |\zeta_i(t')|^2 + (1-\eta_i)|\zeta_i(t_r)|^2 = |c_{e,i}|^2$ and $\eta_i^2 |\zeta_{ii}(t',t'')|^2 + (1-\eta_i)\eta_i|\zeta_{ii}(t',t_r)|^2 + (1-\eta_i)\eta_i|\zeta_{ii}(t_r,t')|^2 +(1-\eta_i)^2|\zeta_{ii}(t_r,t_r)|^2= |c_{ee,i}|^2$ to simplify the matrix elements.\\

\noindent\textbf{Noise counts} In case of a false heralding event by a noise count, the average density matrix consists of two parts $\rho_0$ and $\rho_{\text{lost}}$
\begin{equation}
\rho_{\text{noise}} = p_d (\rho_{0} + \rho_{\text{lost}})
\end{equation}
$\rho_{0}$ corresponds to the situation where no photon is emitted and $\rho_{\text{lost}}$ where all emitted photons are lost. $\rho_0$ is given by
\begin{equation}
\begin{split}
\rho_{0} =& \ket{\Psi_{4,0}}_A \bra{\Psi_{4,0}}_A \otimes \ket{\Psi_{4,0}}_B \bra{\Psi_{4,0}}_B\\
=& 
\begin{pmatrix}
a_{00} & a_{01}  \\
a_{01}^\ast & a_{11} \\
\end{pmatrix}
\otimes
\begin{pmatrix}
b_{00} & b_{01}  \\
b_{01}^\ast & b_{11} \\
\end{pmatrix}
\end{split}
\end{equation}
with elements 
\begin{align}
\begin{split}
a_{00} = \alpha_A c^2_{0,A}
\end{split}
\\
\begin{split}
a_{01} = \saa \smaa c_{0,A} e^{i\vartheta_A}
\end{split}
\\
\begin{split}
a_{11} = (1-\alpha_A)
\end{split}
\end{align}
and similar elements for $b_{ij}$. $\rho_{r}$ represents the density matrix in case all photons are lost
\begin{equation}
\begin{split}
\rho_{\text{lost}} =& \sum \ket{\Psi_{4,0,r}}\bra{\Psi_{4,0,r}}  \\
=& 
\begin{pmatrix}
a_{00} & 0 & 0 & 0 \\
0 & a_{11}  & 0  & 0 \\
0 & 0& a_{22} & 0 \\
0 & 0 & 0 & 0 \\
\end{pmatrix},
\end{split}
\end{equation}
with elements 
\begin{align}
\begin{split}
a_{00} = & \alpha_A \alpha_B ( c^2_{0,A} ( (1-\eta_B) |\zeb(t_r)|^2 + (1-\eta_B)^2 |\zebb(t_r, t_r)|^2)  \\
& +   c^2_{0,B} ( (1-\eta_A) |\zea(t_r)|^2 + (1-\eta_A)^2 |\zeaa(t_r, t_r)|^2) \\
& + (1-\eta_A)(1-\eta_B)|\zea(t_r)|^2|\zeb(t_r)|^2)\\
\end{split}
\\
\begin{split}
a_{11} = \alpha_A(1-\alpha_B) ( (1-\eta_A) |\zea(t_r)|^2 + (1-\eta_A)^2 |\zeaa(t_r, t_r)|^2)\\
\end{split}
\\
\begin{split}
a_{22} = (1-\alpha_A)\alpha_B ( (1-\eta_B) |\zeb(t_r)|^2 + (1-\eta_B)^2 |\zebb(t_r, t_r)|^2).\\
\end{split}
\end{align}

\renewcommand{\thefigure}{B\arabic{figure}}
\renewcommand{\theequation}{B.\arabic{equation}}
\renewcommand{\thetable}{B\arabic{table}}
\setcounter{figure}{1}
\setcounter{equation}{1}
\setcounter{table}{1}

\section{Experimental parameters for simulations}\label{app:exp_params}
In Figures \ref{fig:brightness_imbalance} and \ref{fig:photon_dist} we provide simulations for the average fidelity with respect to the maximally entangled states using the model developed in Section \ref{sec:tailored_model}. The parameters we use for these simulations are listed in Table \ref{tab:exp_params}, both for the AB and BC links. 

\begin{table}[b]
	\caption{\textbf{Table with experimental parameters.} The start of the detection window is with respect to the maximum intensity of the optical pulse and the detection probability for each setup is integrated over the detection window. Furthermore, these experiments used the same optical pulse as used in References \cite{Pompili2021,Hermans2021} and is different from the optical pulse shape indicated in Figure \ref{fig:double_exc}a. References \cite{Pompili2021,Hermans2021} use a different arbitrary waveform generator (AWG), the \textit{Tektronix 5014}, which causes the double excitation probabilities of those specific optical excitation pulses to be larger. }
	\label{tab:exp_params}
	\begin{tabular}{|l|r|r|}
		\hline
		& AB & BC \\
		\hline
		Excited state lifetime &12.4 ns&12.4 ns \\
		Detection window start &4 ns&5 ns\\
		Detection window duration  &15 ns&15 ns\\
		Bright state population $\alpha_A$ & 0.07&- \\
		Bright state population $\alpha_B$ & 0.05&0.05 \\
		Bright state population $\alpha_C$ & -&0.1 \\
		Detection probability $\eta_A \int\mathcal{E}^2 dt$ &3.8e-4&-\\
		Detection probability $\eta_B \int\mathcal{E}^2 dt$ &5.2e-4&4.6e-4\\
		Detection probability $\eta_C \int\mathcal{E}^2 dt$ &-&2.8e-4\\
		Double excitation probability &0.06 & 0.08 \\
		Noise count rate &10 Hz&30 Hz\\
		Phase stability $\sigma_{\delta\varphi}$  &$30^o$&$21^o$ \\
		FWHM frequency difference $\Delta f_{\text{FWHM}}$ &13 MHz &13 MHz \\
		Polarization mismatch &$ 8^o $&$ 8^o $\\	
		\hline
	\end{tabular}
\end{table}

\end{appendices}

\end{document}